\documentclass[fleqn,10pt]{wlscirep}

\usepackage[utf8]{inputenc}
\usepackage[T1]{fontenc}
\usepackage{lineno}
\usepackage{multirow}
\usepackage{longtable} 
\usepackage{xcolor}
\definecolor{darkorange}{rgb}{0.8, 0.4, 0.0} 
\definecolor{darkred}{rgb}{0.35, 0.0, 0.0}
\usepackage{colortbl}
\usepackage{array}
\usepackage{hyperref}
\usepackage{makecell} 

\usepackage{nicematrix}

\title{Spatially Disaggregated Energy Consumption and Emissions in End-use Sectors for Germany and Spain}

\author[1,2]{Shruthi Patil}
\author[1,*]{Noah Pflugradt}
\author[1]{Jann M. Weinand}
\author[3,2]{Jürgen Kropp}
\author[1,4]{Detlef Stolten}

\affil[1]{Jülich Systems Analysis, Forschungszentrum Jülich, Wilhelm-Johnen-Straße, 52425 Jülich, NRW, Germany}
\affil[2]{University of Potsdam, Institute for Environmental Science and Geography, Karl-Liebknecht-Str. 24-25, Potsdam-Golm, 14476 Brandenburg, Germany}
\affil[3]{Potsdam Institute for Climate Impact Research (PIK), Member of the Leibniz Association, P.O. Box 60 12 03, Potsdam D-14412, Brandenburg, Germany}
\affil[4]{Chair for Fuel Cells, RWTH Aachen University, c/o Jülich Systems Analysis, Forschungszentrum Jülich, 52425 Jülich, NRW, Germany}

\affil[*]{corresponding author: Noah Pflugradt (n.pflugradt@fz-juelich.de)}

\begin{abstract}
High-resolution energy consumption and emissions datasets are essential for localized policy-making, resource optimization, and climate action planning. They enable municipalities to monitor mitigation strategies and foster engagement among governments, businesses, and communities. However, smaller municipalities often face data limitations that hinder tailored climate strategies. This study generates detailed final energy consumption and emissions data at the local administrative level 
for Germany and Spain. Using national datasets, we apply spatial disaggregation techniques with open  
data sources. A key innovation is the application of XGBoost for imputing missing data, combined with a 
stepwise spatial disaggregation process incorporating district- and province-level statistics. 
Prioritizing reproducibility, our open-data approach provides a scalable framework for 
municipalities to develop actionable climate plans. To ensure transparency, we assess the reliability of imputed values and assign confidence ratings to the disaggregated data.

\end{abstract}
\begin{document}

\flushbottom
\maketitle

\thispagestyle{empty}


\section*{Background \& Summary}

Spatially and sectorally detailed energy consumption and emissions data is a valuable resource for the energy 
and climate research community. For instance, the energy consumption in road transport can support the planning of electric vehicle 
charging infrastructure, facilitating the transition to electric mobility and assessing its impact on grid load \cite{deb2018review}. 
However, bottom-up inventories are rare at the municipal level and often omit cross-regional traffic \cite{netzerocities-valencia}, leaving a considerable share of national final energy consumption unaccounted for in local inventories.

Such granular data is also critical for formulating localized climate strategies, including Sustainable Energy 
and Climate Action Plans (SECAPs) led by the Covenant of Mayors \cite{kona2018covenant}. These plans include a Final Energy Consumption (FEC) 
stock-taking across various sectors, including residential, Emissions Trading System (ETS) and non-ETS industries, and public and private transport. These datasets are particularly valuable for municipalities with limited statistical data infrastructure or expertise in compiling sectoral baseline energy consumption and emissions reports. 

Currently, sectorally-detailed emissions and FEC data are unavailable in official databases at fine-scale spatial resolutions, such as the Local Administrative Unit (LAU) level. However, various open data sources provide complementary information, such as land use, land cover, population, and industrial site locations. By leveraging these open databases, it is possible to estimate the spatial distribution of emissions and FEC for individual sub-sectors and allocate national-level values proportionally across finer spatial scales. 

The spatial disaggregation of sectoral emissions and FEC has been explored in previous studies \cite{crippa2018gridded, moran2021estimating, valencia2022downscaling, risch2024scaling}. However, the spatial resolution of existing data products varies, often aligning with federal states or district levels. Additionally, past efforts frequently overlook detailed sub-sectors, leading to gaps in the representation of specific activities and industries. For example, emissions in non-energy-intensive activities such as construction and food, beverages, and tobacco manufacturing are not included in the previous works.

By providing detailed, spatially disaggregated energy data, this data descriptor can support municipalities in fulfilling inventory preparation and reporting requirements. This study spatially disaggregates FEC and greenhouse gas (GHG) emissions for end-use sectors \textemdash industry, 
transport, agriculture, households, and commerce \textemdash from the national level to the LAU level. It incorporates detailed sub-sectors such as rail transport, road transport by cars, motorcycles, and various industries, including iron and steel, non-ferrous metals, and textile and leather industries, wherever sector-specific national data is available.

This study focuses on Germany and Spain. Germany is selected as it is the largest GHG emitter in the EU \cite{european_commission_joint_research_centre_ghg_2024}, while Spain is chosen due to its national government's emphasis on the role of local authorities in meeting climate targets outlined in its national climate action plans \cite{necp_spain}. In this context, top-down FEC and emissions inventories can support Spanish local governments in developing climate action plans that align with national strategies.

\section*{Methods}

\subsection*{Overview}
\begin{figure}[h]
    \centering
    \includegraphics[width=\linewidth]{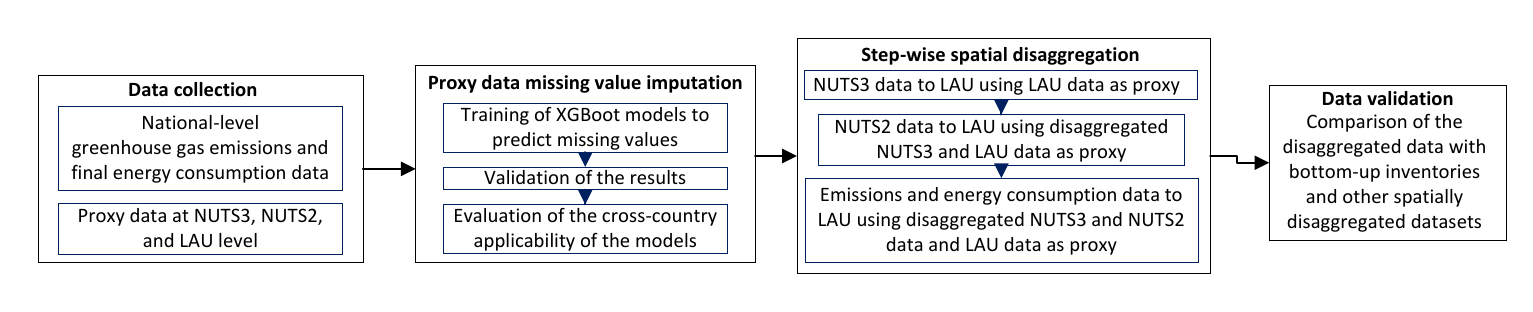}
    \caption{The steps involved in the spatial disaggregation of emissions and Final Energy Consumption (FEC) data 
    from country (NUTS0) to municipal (LAU) level.}
    \label{fig:workflow_no_data_sources}
\end{figure}

The spatial disaggregation of FEC and emissions data is carried out in four steps, as illustrated in Figure \ref{fig:workflow_no_data_sources}. The following paragraphs provide an overview of these four steps, before going into more detail 
in the sub-sections of this Methods section.

\textbf{1. Data collection.} FEC and emissions data for end-use sectors are available at the national level from Eurostat \cite{eurostatmainpage}. This serves as the target
data for this work, which is disaggregated to the municipal level. For the disaggregation, a wide range of methods exists in the literature, from 
simple techniques based on area proportions to more advanced approaches based on machine learning or geostatistical methods 
\cite{PATIL2024100386}. 

\begin{figure}[h]
    \centering
    \includegraphics[width=\linewidth]{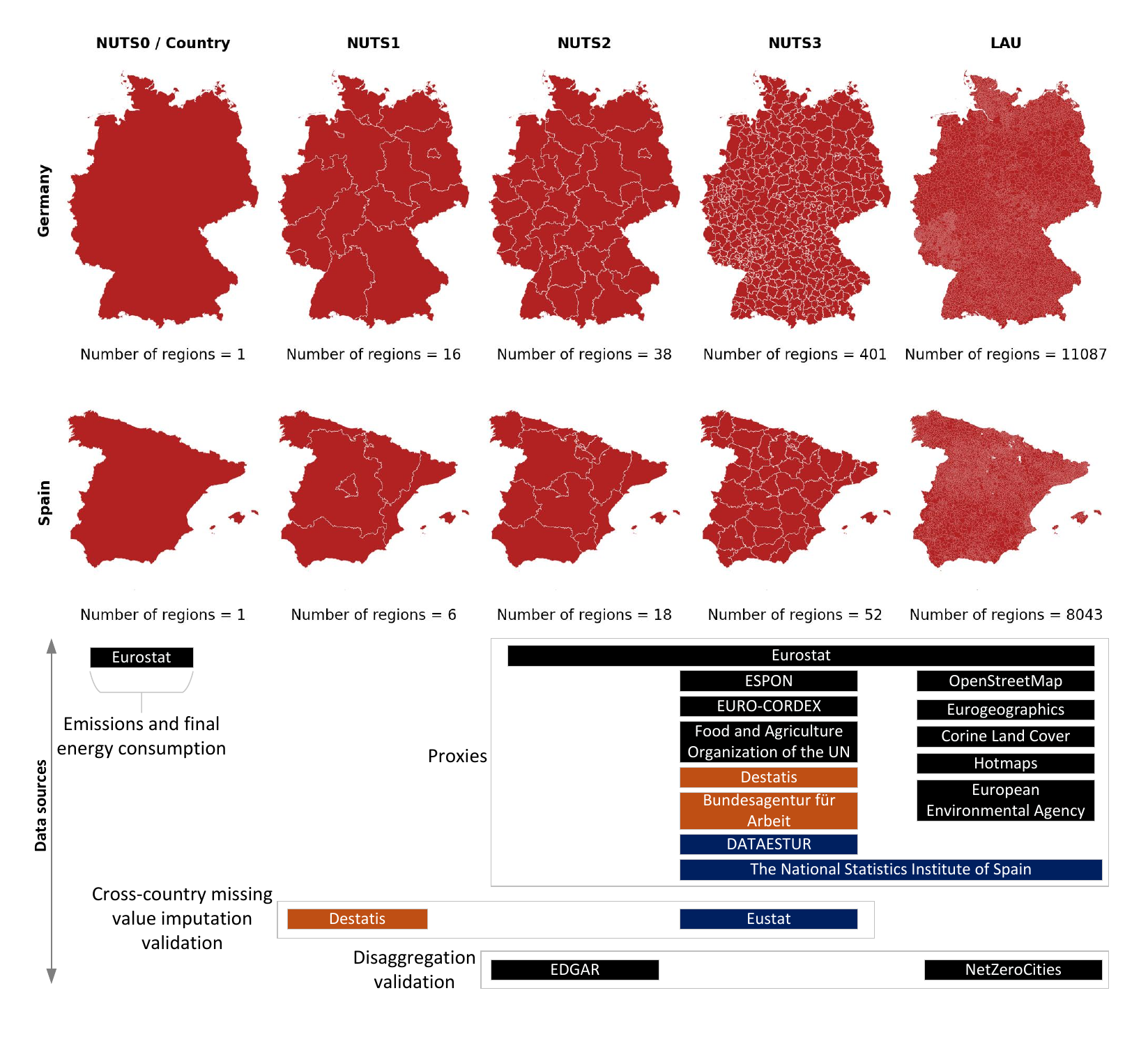}
    \caption{The spatial hierarchy in Germany and Spain, showing the availability of various proxy datasets from public data sources at 
    different spatial levels. The data sources highlighted in orange and blue provide data only for Germany and Spian, respectively.
    Proxy data undergoes a stepwise spatial disaggregation to achieve final proxies at the LAU level. Emissions 
    and FEC data, available at the NUTS0 level from Eurostat, is then disaggregated to LAU based on these final proxies.}
    \label{fig:spatial_hierarchy_with_data_sources}
\end{figure}

This work adopts a proxy-based approach, leveraging domain knowledge to identify the most relevant spatial proxies 
for the target data. The disaggregation process involves allocating national-level values to municipalities based on the 
proportion of the selected proxy in each region. For example, residential building emissions at the national level can be spatially distributed to municipalities based 
on their respective population shares, as population density is a key determinant of residential emissions. Similarly, 
different end-use sectoral data are disaggregated using suitable proxies. These proxies are obtained from publicly 
available databases, including Eurostat, Corine Land Cover \cite{corine_land_cover} and OpenStreetMap \cite{open_street_map}. A comprehensive overview of all data sources is provided in Figure \ref{fig:spatial_hierarchy_with_data_sources}. 

The FEC and emissions data were collected for the year 2022. Proxy data, on the other hand, spans multiple years, as certain datasets\textemdash such as land use and land cover \textemdash are not updated annually. However, all proxy datasets used represent the most recent year for which data was available at the time of collection.

\textbf{2. Missing value imputation.} The collected proxy data may have missing values. To address this issue, we employ the XGBoost algorithm for imputing missing values \cite{chen2016xgboost}. This step ensures that the dataset is complete for the subsequent spatial disaggregation process. The missing values primarily occur in country-specific datasets. To assess the model's performance, we conduct two evaluations: (1) assessing the predictive accuracy within a country by setting aside a portion of the data for validation, and (2) evaluating the model's predictive capacity in a country where the dataset is entirely absent. The latter is achieved by leveraging available data at intermediate spatial levels, such as states, within the country. The data sources employed in this cross-country missing value imputation evaluation are listed in Figure \ref{fig:spatial_hierarchy_with_data_sources}. 

\textbf{3. Stepwise spatial disaggregation.} Figure \ref{fig:spatial_hierarchy_with_data_sources} depicts the Nomenclature des Unités Territoriales Statistiques (NUTS) spatial hierarchy in Germany and Spain. Starting from the highest level at the country (NUTS0), each subsequent level (NUTS1, NUTS2, NUTS3) divides the regions into smaller areas, enhancing spatial resolution. LAU regions further subdivide the NUTS3 regions, forming the finest spatial resolution for administrative regions. Figure \ref{fig:spatial_hierarchy_with_data_sources} shows that only some proxy data is readily available at the LAU level, while most statistical data is typically available at NUTS3 or NUTS2 levels. Additionally, most proxy datasets do not cover the Canary Islands in Spain. Due to this limitation, these regions are excluded from the scope of this study.

In this work, we perform a stepwise spatial disaggregation. Initially, proxy data available at the NUTS3 level is disaggregated using LAU proxy data, to achieve finer resolution. Subsequently, NUTS2 proxy data is disaggregated to the LAU level using both the LAU data and the previously disaggregated NUTS3 data as proxies. Finally, the emissions and FEC data are disaggregated to the LAU level. This approach improves the accuracy of the disaggregated data by progressively refining estimates.

\textbf{4. Data validation.} Finally, the spatial disaggregation results are compared with the values reported in local inventories (NetZeroCities \cite{noauthor_netzerocities_nodate}) and  an open-source sub-national emissions dataset (EDGAR \cite{crippa2024insights}), ensuring alignment with the sub-sectors considered in this study. The details of this validation process are presented in the technical validation section of this manuscript.

\subsection*{Data Collection}

For the disaggregation work, this study utilizes three types of data: (i) FEC data for various sub-sectors, at the national level, (ii) emissions data for various sub-sectors also at the national level and (iii) proxy data relevant to each sub-sector, which facilitates the disaggregation of FEC and emissions data from the national to the LAU level. The details regarding the collection of each dataset are discussed in the following sections.

\subsubsection*{Final Energy Consumption}
\begin{figure}[h]
    \centering
    \includegraphics[width=\linewidth]{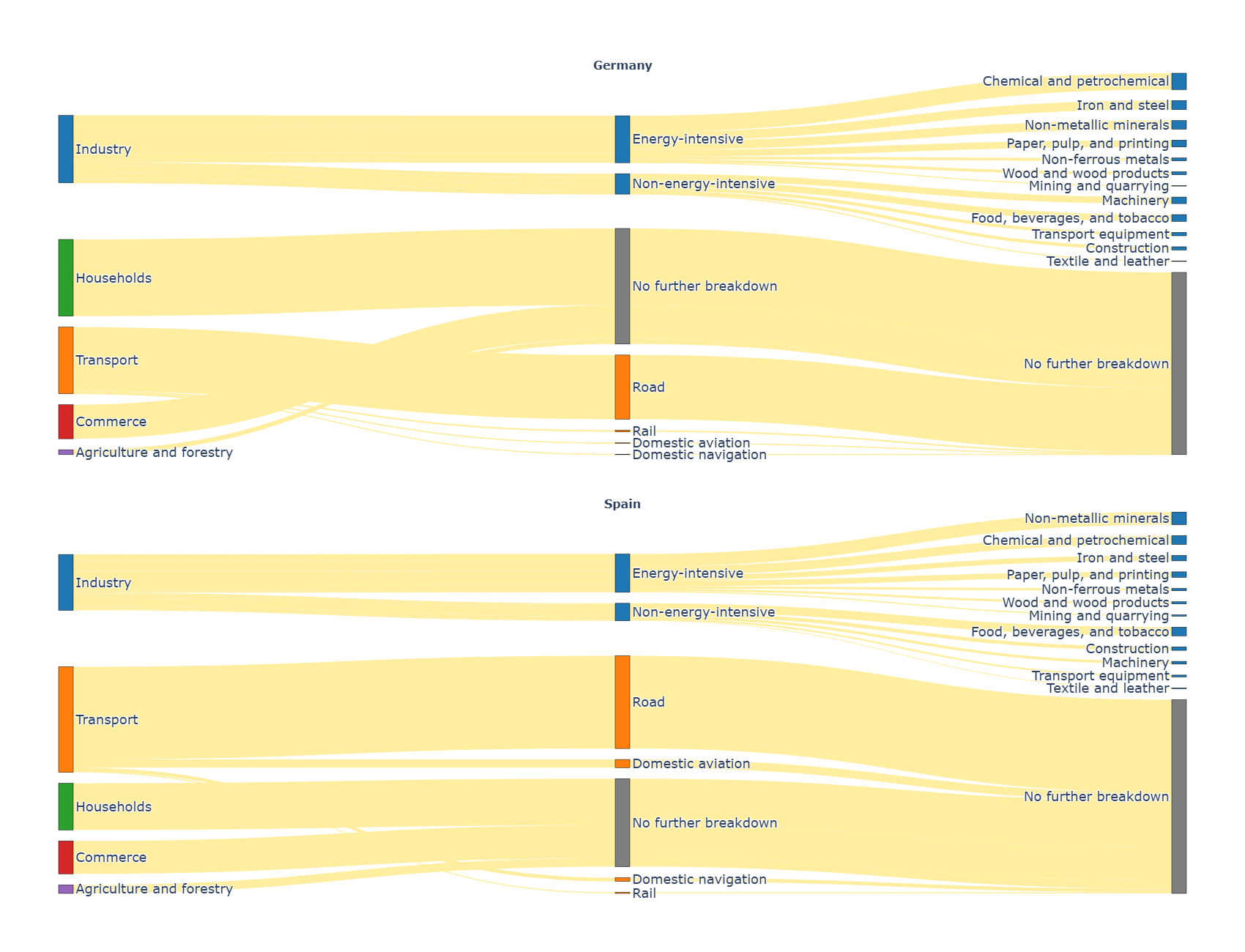}
    \caption{Breakdown of end-use FEC sectors as reported in Eurostat, with Germany at the top and 
    Spain at the bottom.}
    \label{fig:fec_categorisation}
\end{figure}

The FEC data, reported at the national level, is imported from the energy balance sheet published on Eurostat \cite{eurostat_complete_2023}.
While the FEC data for both energy and non-energy use is reported on Eurostat, only the energy use FEC is considered here. The breakdown of the end-use sectors for emissions reporting on Eurostat is shown in Figure \ref{fig:fec_categorisation}. 

The industry sector is broken down into energy-intensive and non-energy-intensive industries. Energy-intensive industries include iron and steel, chemical and petrochemical, non-ferrous metals, non-metallic minerals, mining and quarrying, paper, pulp, and printing, and wood and wood products manufacturing industries \cite{iea2009energy}. Non-energy-intensive industries include transport equipment, machinery, food, beverages, and tobacco, textile and leather manufacturing industries, construction, and other industries that are not specified 
elsewhere \cite{iea2009energy}. A similar categorisation is provided by Eurostat. 

The transport sector is categorized into rail, road, domestic aviation, and domestic navigation. Additionally, the commerce and agriculture and forestry sectors are included, though they are not further broken down.

\subsubsection*{Greenhouse Gas Emissions}
\begin{figure}[h]
    \centering
    \includegraphics[width=\linewidth]{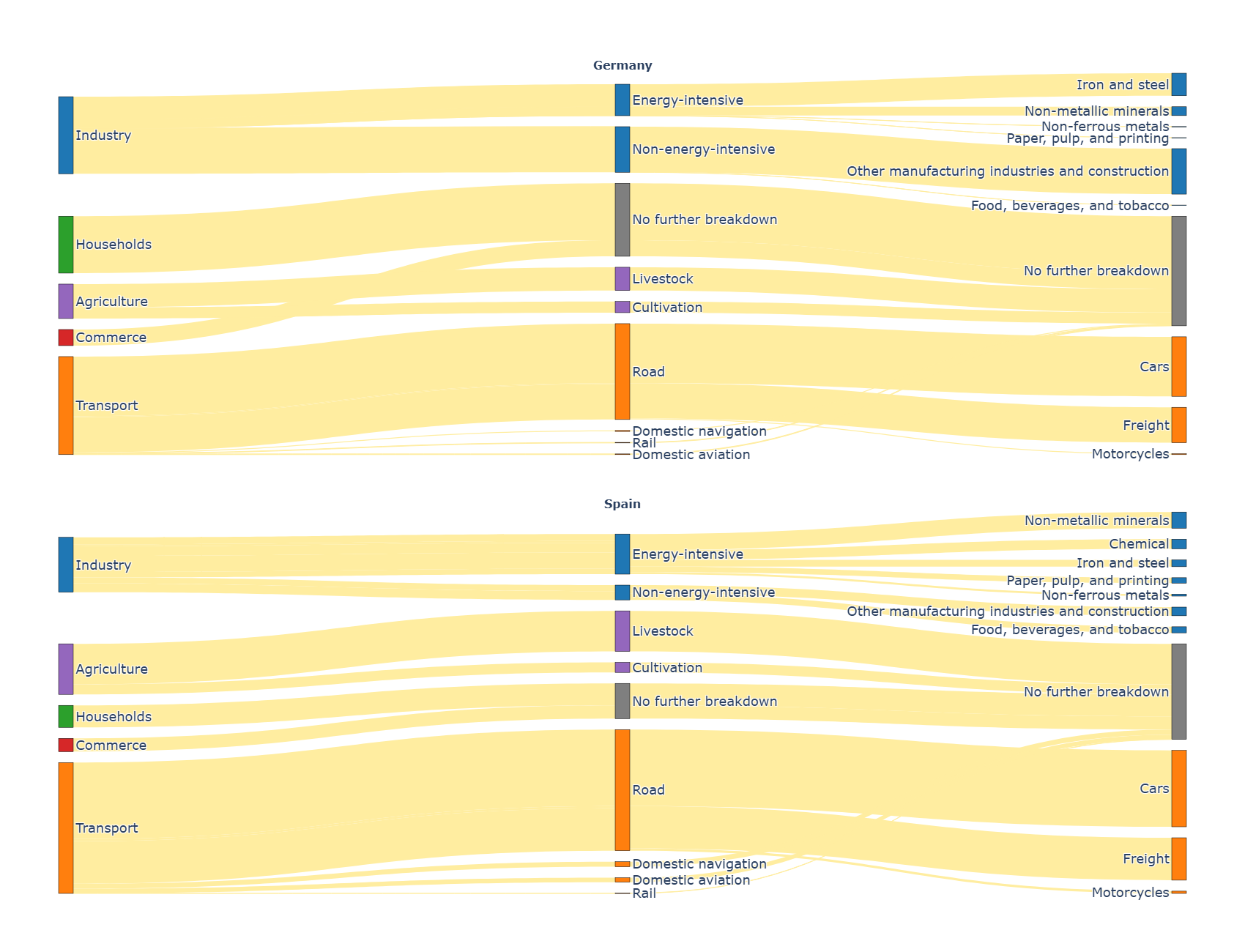}
    \caption{Breakdown of end-use emission sectors as reported in Eurostat, with Germany at the top and 
    Spain at the bottom. Note: Emissions from the chemical industry are not reported for Germany on 
    Eurostat and are therefore absent from the figure. Consequently, emissions from energy-intensive industries appear 
    lower than those from non-energy-intensive industries.}
    \label{fig:emissions_categorisation}
\end{figure}

The GHG emissions data used in this study is sourced from Eurostat \cite{eurostat_greenhouse}. Figure \ref{fig:emissions_categorisation} illustrates how end-use sectors are categorized for emissions reporting by Eurostat. Compared to the FEC sector classification, the industry sector emissions data provided by Eurostat is less detailed. Notably, emissions from the chemical industry are not reported for Germany, and as such, these emissions are not disaggregated in this study.

In contrast, the transport sector is more granular in the Eurostat data compared to the FEC categorisation. Here, sub-sectors such as rail, road, domestic aviation, and domestic navigation, are further broken down into categories like light-duty trucks, heavy-duty trucks and buses, cars, and motorcycles. For the purposes of this study, emissions from light-duty trucks and heavy-duty trucks and buses are grouped under freight transport, as specific proxies for these vehicle types are not available in publicly accessible datasets.

It is important to highlight that GHG emissions in the sectors examined here primarily stem from fuel combustion. Process-related emissions in the industrial sector are excluded, with one exception: the agriculture sector, which includes non-combustion emissions. Agricultural emissions are considered under two categories—livestock and cultivation. Although Eurostat provides further disaggregation into subcategories such as enteric fermentation, manure management, agricultural soil management, and crop residue burning, only livestock and cultivation are included in this study due to the absence of matching proxies in open-source data.

\subsubsection*{Proxy Data}
Proxy data serves as the foundation for spatially disaggregating emissions and FEC across different end-use sectors. The 
selection of proxy data was guided by the following criteria:
\begin{enumerate}
    \item \textbf{Availability in open databases}: The data must be accessible through publicly available databases to enhance 
    transparency and reproducibility.

    \item \textbf{Regional resolution}: The data should be available at a regional scale, such as NUTS1, NUTS2, NUTS3, or LAU, 
    to facilitate the downscaling of emissions and FEC data reported at NUTS0.

    \item \textbf{Relevance to end-use sectors}: The selected data must be relevant to at least one of the end-use sectors analyzed 
    in this study. For instance, population data can be used to disaggregate household-related emissions and FEC. Heating degree days help refine the spatial distribution of FEC by accounting for higher heat demand in colder regions. In addition, industrial locations and employment data provide insights into the spatial distribution of industrial emissions and FEC. Furthermore, vehicle stock data supports the disaggregation of road transport emissions and energy consumption. Finally, land use and land cover classifications (e.g., rice fields, vineyards) assist in distributing cultivation-related FEC and emissions.
    
    \item \textbf{Data completeness}: The dataset should have minimal missing values to ensure a robust analysis. Here, less than 20\% missing data was preferred.
\end{enumerate}

Subsequently, an overview of the collected proxy data at different spatial levels is provided, beginning with the LAU level.

\textbf{Proxy data at LAU level.} The data collected at LAU level in Germany and Spain is summarized in Tables \ref{tab:lau_proxy_data_1} and \ref{tab:lau_proxy_data_2}. This dataset includes general statistics such as population and area, sourced from Eurostat, 
which are directly available for each LAU region. Land use and land cover data, including areas such as continuous urban fabric, are derived from the Corine Land Cover database. This database provides raster data at a fine-scale resolution of 100 square meters, allowing for spatial overlap and aggregation of land cover classes per LAU region. Similarly, air pollution data is obtained from the European Environment Agency, which provides data at a spatial resolution of 1 square kilometer. In addition, Eurogeographics provides data on the railway network, while OpenStreetMap offers information on road networks and building counts. These vector datasets are overlaid with LAU regions to extract aggregates per region. 

Data on industrial sites was available in three databases: sEEnergies \cite{seenergies_open_data}, Global Steel Plant Tracker 
\cite{global_steel_plant_tracker}, and Hotmaps. To determine the most suitable source, the datasets were compared for their level 
of detail and coverage.

\begin{itemize}
    \item \textbf{sEEnergies} provides industrial site locations along with fuel and electricity demand information for industries 
    such as iron and steel, chemicals, non-ferrous metals, non-metallic minerals, paper and printing, and refineries. 

    \item \textbf{Global Steel Plant Tracker} focuses solely on iron and steel plant locations, annotating them with energy 
    demand and employment data, though many sites lack complete information. 

    \item  \textbf{Hotmaps} includes locations for cement and glass industries in addition to those covered by sEEnergies, with 
    emissions data provided for each site, though data is missing for many locations.
\end{itemize}

\begin{figure}[h]
    \centering
    \includegraphics[width=\linewidth]{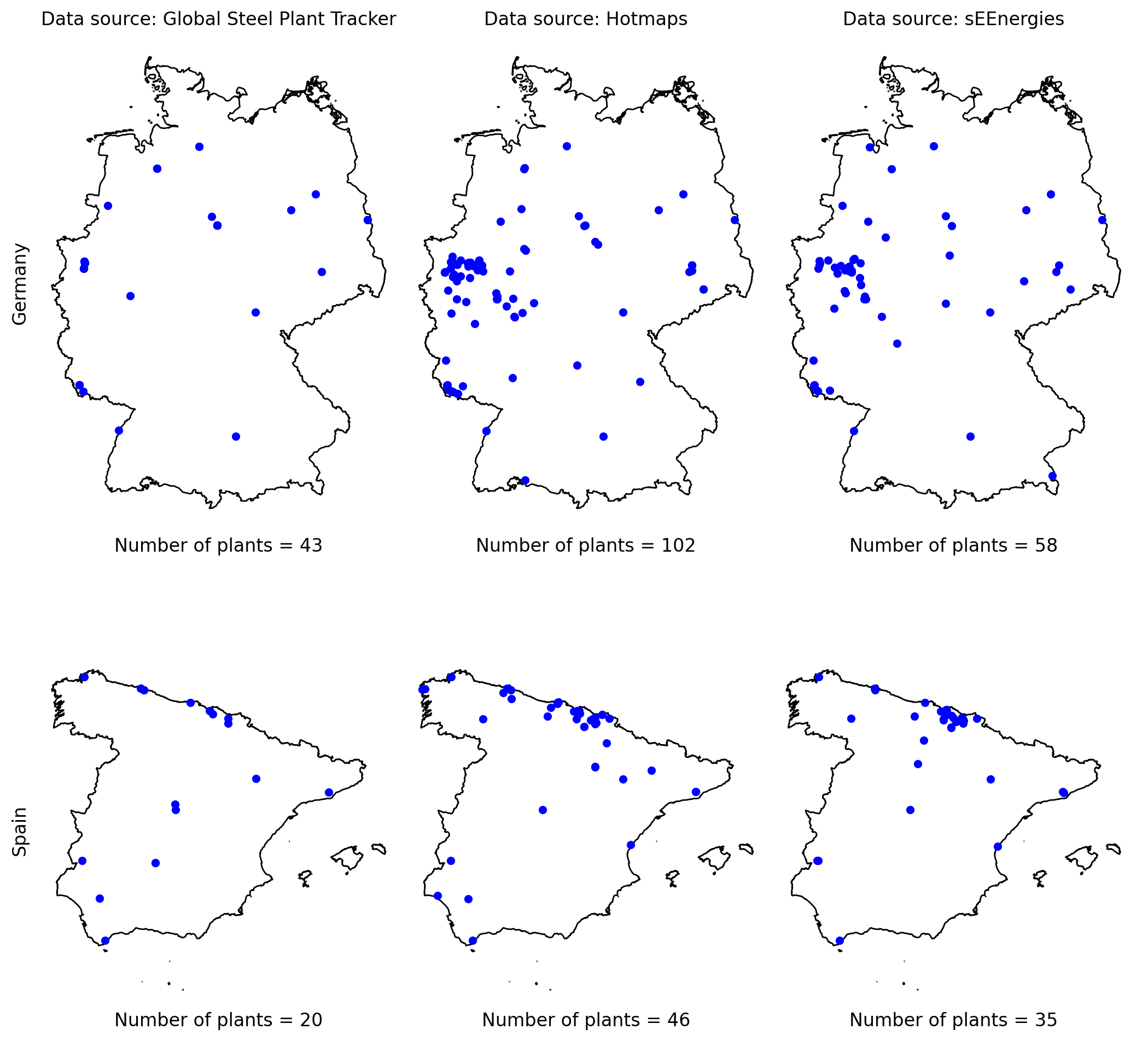}
    \caption{The distribution and number of iron and steel industries as reported by three open databases: Global Steel Plant Tracker, 
    Hotmaps, and sEEnergies. The figure highlights the differences in coverage among these sources, with Hotmaps providing the most 
    comprehensive dataset.}
    \label{fig:iron_and_steel_plants}
\end{figure}

A comparative analysis was performed to select the most comprehensive source for each sector. For the iron and steel industry, 
a comparison of site counts across the three datasets was conducted, as illustrated in Figure \ref{fig:iron_and_steel_plants}. 
For other industries, a comparison between sEEnergies and Hotmaps was performed (see Table \ref{tab:industrial_sites_comparison}). Hotmaps reports a higher number of industrial sites across most categories, except for 
paper and printing industries in Germany, where sEEnergies provides higher counts. Based on this analysis, the Hotmaps database 
was selected as the primary source for obtaining LAU-level industry data, ensuring comprehensive coverage and consistency across sub-sectors.

\textbf{Proxy data at NUTS3 level.} Table \ref{tab:nuts3_proxy_data} provides an overview of the data collected for the German and Spanish NUTS3 regions. Basic statistical information, such as employment and gross domestic product, is published by Eurostat at this spatial level. Heating and cooling degree days and livestock population datasets are available in raster format. With a resolution of approximately 10 square kilometers, these datasets align well with NUTS3 regions, enabling spatial overlap and aggregation at the NUTS3 level. Additionally, some datasets are available only in one of the two countries \textemdash for example, sectorally detailed employment data from the Federal Employment Agency in Germany and company counts from the National Statistics Institute of Spain. 

\textbf{Proxy data at NUTS2 level.} Table \ref{tab:nuts2_proxy_data} summarizes the data available for the German and Spanish NUTS2 regions. All the datasets collected at this level are sourced from Eurostat. 

\subsection*{Missing Value Imputation}

Table \ref{tab:missing_values} provides an overview of the number and percentage of missing values identified in the collected 
proxy data. These gaps are primarily found in datasets that are available only for either Germany or Spain, often due to strict 
data protection regulations preventing certain regions from reporting data. Consequently, missing values must be imputed using 
relevant statistical indicators, such as land use and land cover data when estimating the utilized agricultural area.

Imputing these values is a critical step in spatial disaggregation workflows, as complete proxy data is essential for distributing national totals. Since the distribution process is relative, data quality in one region directly influences the accuracy of all others. To impute these missing values a machine learning model, specifically XGBoost, is employed. The training and validation of the model, as well as the evaluation of the missing value prediction across countries is explained in the following.

\textbf{XGBoost model training.} The datasets with missing values at the LAU level are imputed 
by training an XGBoost model using all other LAU-level variables as potential predictors. Before selecting the final predictors, 
two preprocessing steps are performed to eliminate certain variables:
\begin{enumerate}
    \item \textbf{Removal of non-informative predictors:} Any predictors that have the same value across all regions are discarded 
    because they lack predictive capability.
    \item  \textbf{Correlation analysis:} A pairwise correlation among all potential predictors is examined. If two variables exhibit an 
    absolute correlation of 0.9 or higher, only one is retained to prevent over-representation of highly similar variables in the model.
\end{enumerate}

\begin{figure}[h]
    \centering
    \includegraphics[width=\linewidth]{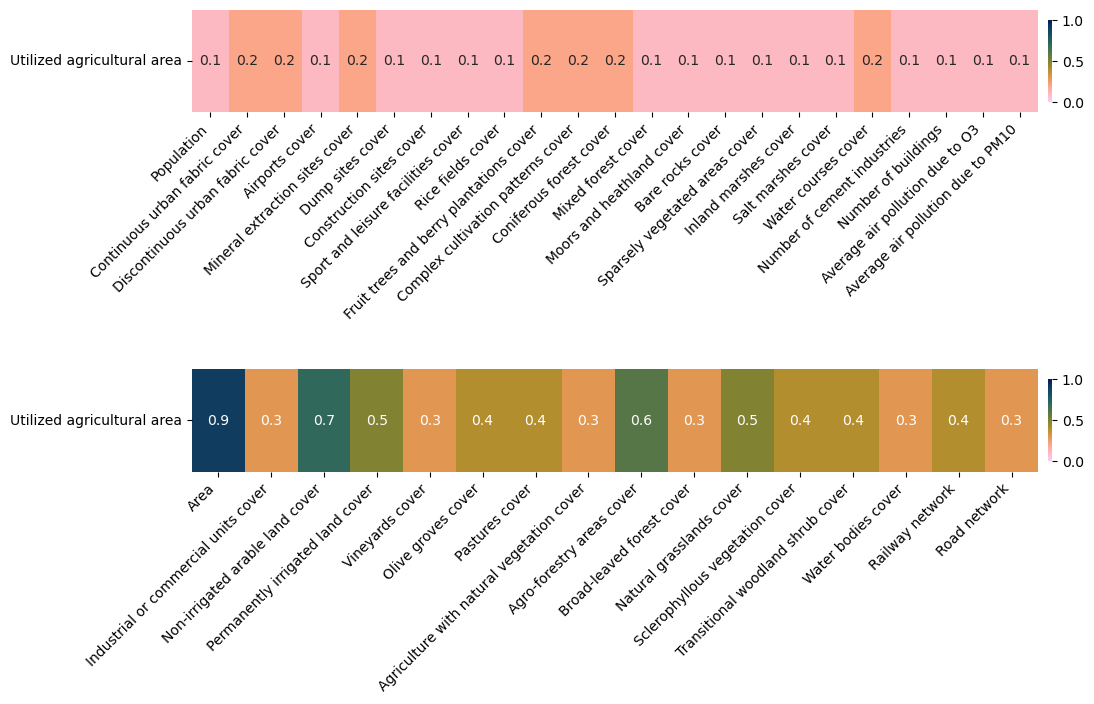}
    \caption{The absolute correlations between utilized agricultural area and different predictors at LAU level. The figure is divided into two sections: the top half displays the least correlated variables, while the bottom half highlights the most correlated ones. For imputing missing values in utilized agricultural area, predictors with correlations of at least 0.1 are used in one set of experiments, while those with correlations of at least 0.5 are considered in another.}
    \label{fig:lau_es_corr}
\end{figure}

The final dataset used as input for the XGBoost model consists of the selected predictors and the variable to be imputed, with only complete records included. Prior to training, 10\% of the data is reserved for model validation, while the remaining 90\% is utilized for two experimental setups. In the first setup, predictors with an absolute Pearson correlation of at least 0.1 with the variable to be imputed are included. In the second setup, the correlation threshold is increased to 0.5. Figure \ref{fig:lau_es_corr} presents the correlations between "utilized agricultural area" and the various predictors used.

In both sets of experiments, hyperparameter tuning is performed using a grid search on the XGBoost model. The hyperparameters tuned include $n\_estimators$, $learning\_rate$, and $max\_depth$, with the model optimized for minimal Root Mean Squared Error (RMSE). To calculate the RMSE, 5-fold cross-validation is applied on the training data, splitting it into five folds. The RMSE is computed for each fold by training the model on four folds and validating on the remaining fold, and the average RMSE across all folds is used as the performance metric for hyperparameter optimization. The final model for data imputation is the one with hyperparameter combination that yields the lowest RMSE.

\begin{figure}[h]
    \centering
    \includegraphics[width=\linewidth]{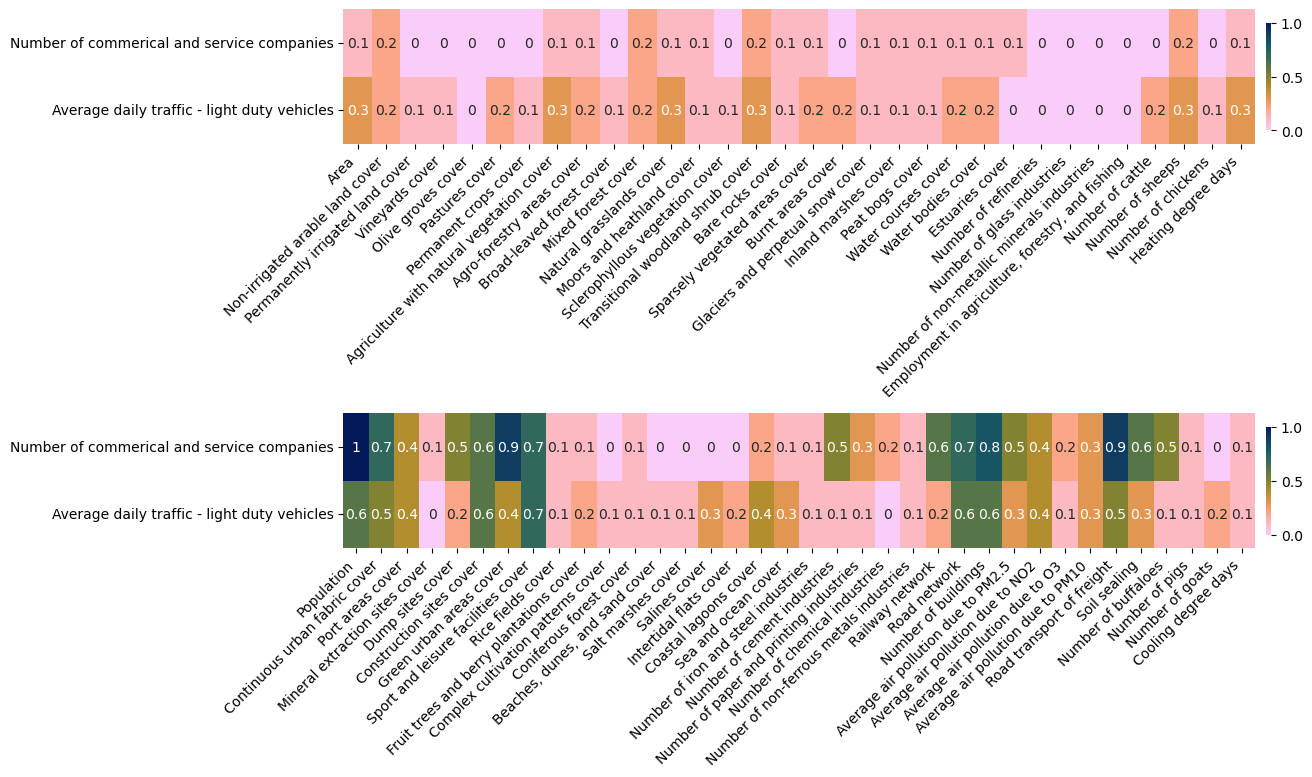}
    \caption{The absolute correlations between number of commercial and service companies and average daily traffic by 
    light duty vehicles, and different predictors at NUTS3 level. The figure is divided into two sections: the top half displays 
    the least correlated variables, while the bottom half highlights the most correlated ones. For imputing missing values, 
    predictors with correlations of at least 0.1 are used in one set of experiments, while those with 
    correlations of at least 0.5 are considered in another.}
    \label{fig:es_nuts3_corr}
\end{figure}

\begin{figure}[h]
    \centering
    \includegraphics[width=\linewidth]{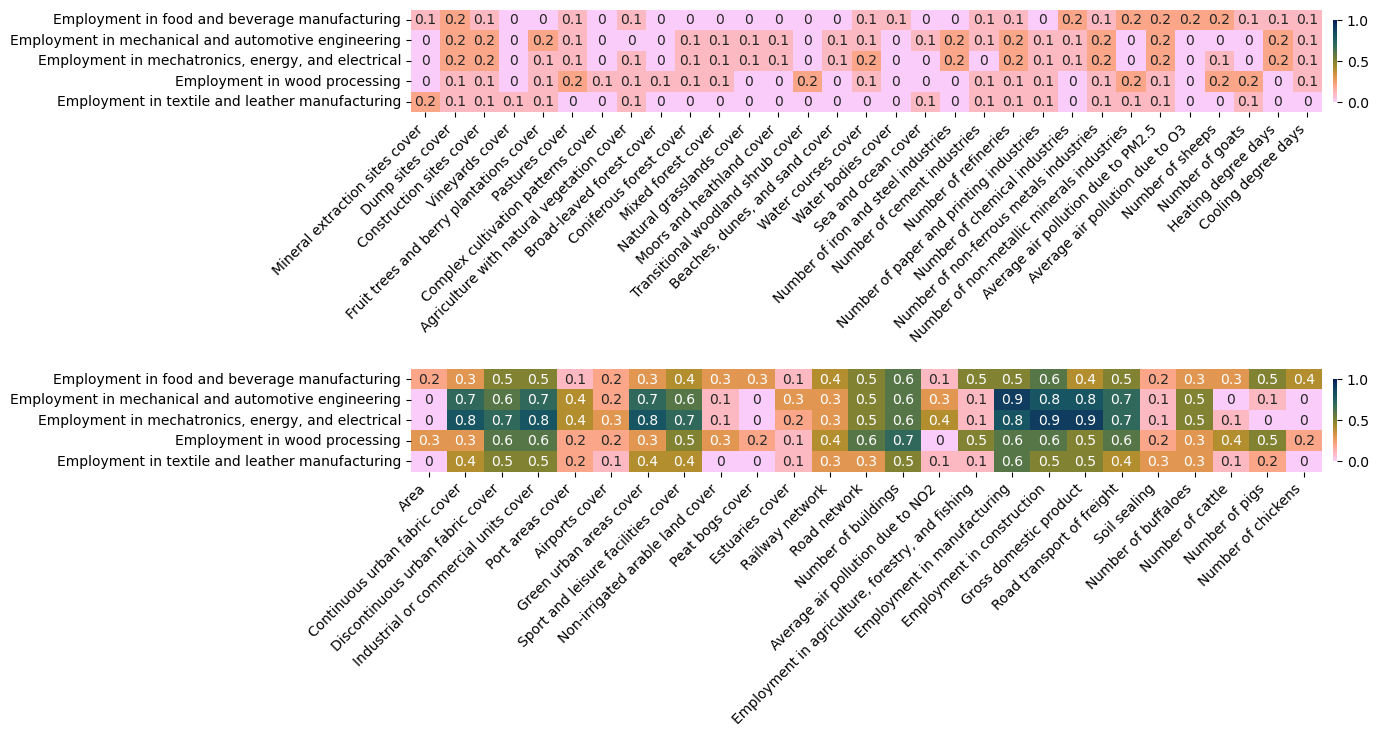}
    \caption{The absolute correlations between employment data and different predictors at NUTS3 level. The figure is divided into 
    two sections: the top half displays the least correlated variables, while the bottom half highlights the most correlated ones.
    For imputing missing values, predictors with correlations of at least 0.1 are used in one set of experiments, while those with 
    correlations of at least 0.5 are considered in another.}
    \label{fig:employment_corr}
\end{figure}

\begin{figure}[h]
    \centering
    \includegraphics[width=\linewidth]{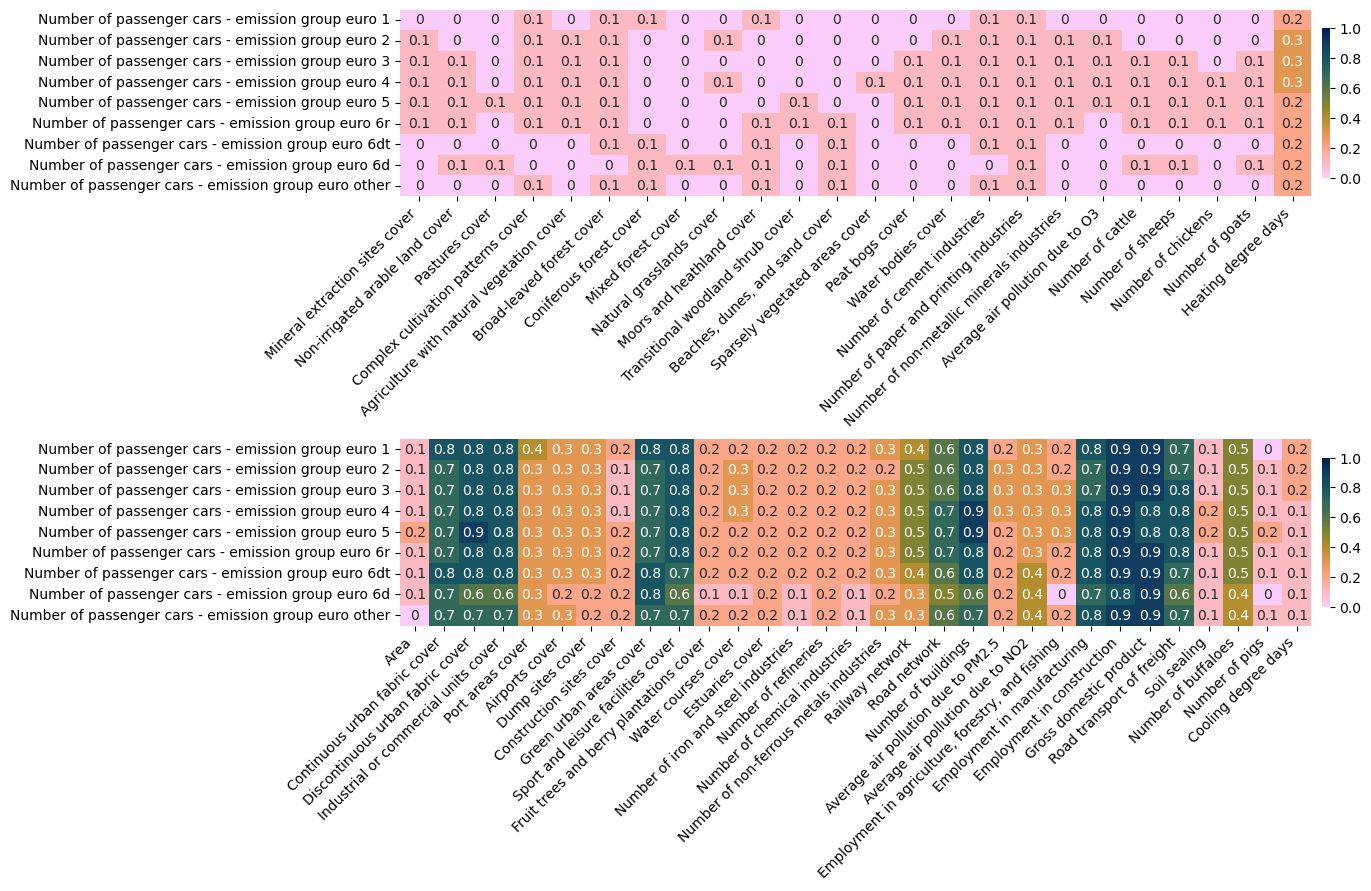}
    \caption{The absolute correlations between the number of passenger cars per emission group and different predictors 
    at NUTS3 level. The figure is divided into two sections: the top half displays the least correlated variables, while the bottom 
    half highlights the most correlated ones. For imputing missing values, predictors with correlations of at least 0.1 are used in 
    one set of experiments, while those with correlations of at least 0.5 are considered in another.}
    \label{fig:passenger_cars_emissions_corr}
\end{figure}

\begin{figure}[h]
    \centering
    \includegraphics[width=\linewidth]{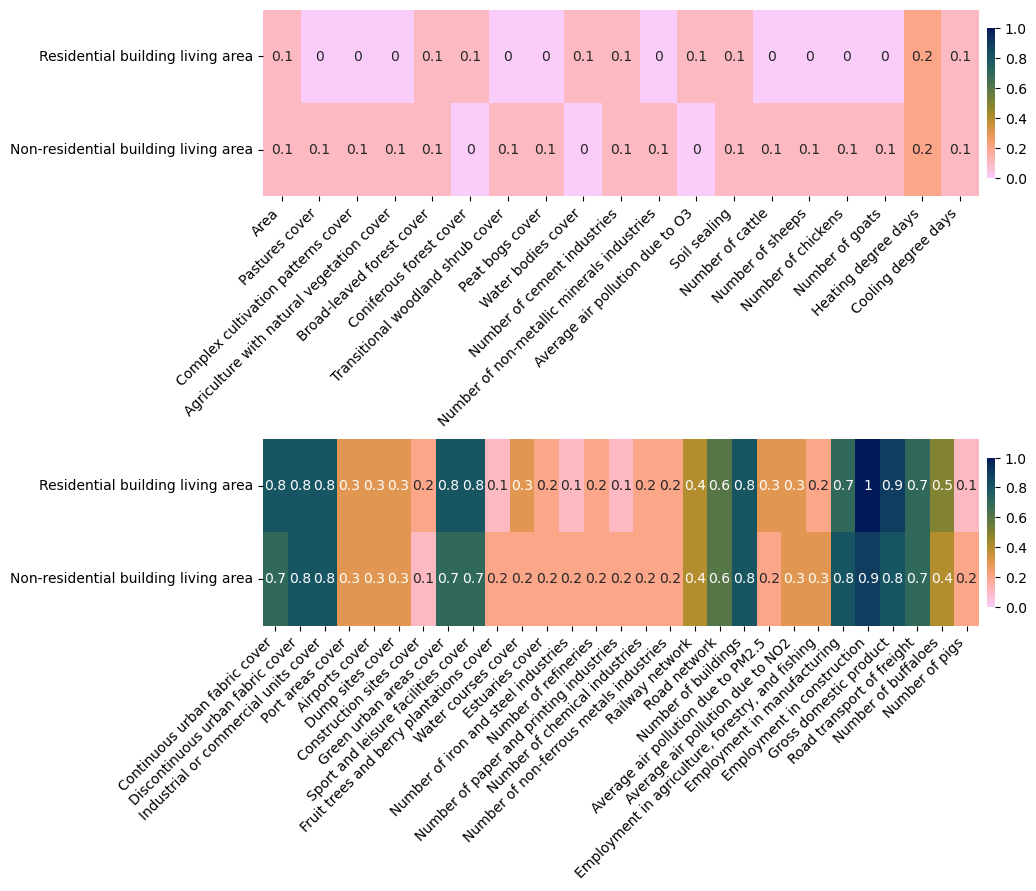}
    \caption{The absolute correlations between the building living area and different predictors at NUTS3 level. The figure is divided into 
    two sections: the top half displays the least correlated variables, while the bottom half highlights the most correlated ones.
    For imputing missing values, predictors with correlations of at least 0.1 are used in one set of experiments, while those with 
    correlations of at least 0.5 are considered in another.}
    \label{fig:building_living_area_corr}
\end{figure}

A similar approach is used in the case of variables with missing values at the NUTS3 level. Here, the potential predictors are all variables at NUTS3 level without missing values, as well as LAU variables with no missing data, aggregated to the NUTS3 level. Figures \ref{fig:es_nuts3_corr}, \ref{fig:employment_corr}, \ref{fig:passenger_cars_emissions_corr}, and 
\ref{fig:building_living_area_corr} illustrate the correlations between the NUTS3 variables with missing values and various predictors.

\textbf{XGBoost model validation.} Table \ref{tab:missing_value_imputation_results} presents the training and validation errors corresponding to the previously discussed correlation thresholds. While a lower RMSE indicates better model performance, its lack of fixed upper or lower 
bounds makes accuracy interpretation challenging. Therefore, the R-squared error is also provided, where values closer to 1 
signify better performance.

To ensure transparency regarding the quality of the generated values in this work, a five-level confidence schema is introduced: 
\textit{VERY HIGH}, \textit{HIGH}, \textit{MEDIUM}, \textit{LOW}, and \textit{VERY LOW}. This labeling system simplifies the 
interpretation of data quality compared to standard error metrics such as RMSE or R-squared values. The confidence assignment begins during the data collection phase, where all non-missing values are automatically labeled with 
\textit{VERY HIGH} confidence. In contrast, missing values are assigned one of the remaining confidence levels (\textit{HIGH}, 
\textit{MEDIUM}, \textit{LOW}, or \textit{VERY LOW}) based on the resulting R-squared scores. Table 
\ref{tab:imputation_confidence_level} defines the threshold mappings between R-squared values and confidence levels for imputed 
data. While the XGBoost model is optimized to minimize RMSE, R-squared scores are used for quality ratings due to their upper limit of 
1, which provides a consistent and interpretable confidence scale across different variables. To clearly distinguish the 
best-performing model and its corresponding confidence level, the color scheme from Table \ref{tab:imputation_confidence_level}
is applied to Table \ref{tab:missing_value_imputation_results}.

The trained XGBoost models effectively predict missing values 
for most variables (see Table \ref{tab:missing_value_imputation_results}). However, the variables "employment in the food and beverage manufacturing sector" have \textit{LOW} prediction quality, 
and "employment in textile and leather manufacturing" is classified as \textit{VERY LOW}. The poor predictions for food and 
beverage manufacturing can be attributed to the lack of relevant predictor data at the NUTS3 level. In addition to this 
limitation, the low prediction quality for textile and leather manufacturing is further exacerbated by a higher proportion 
of missing values, with 34 out of 401 records missing, compared to other datasets at the NUTS3 level. The variable "Average daily traffic - light duty vehicles" shows the poorest prediction results. The R-squared values are negative, 
indicating that none of the predictors contribute meaningfully to the predictions. Out of 52 records, 10 have missing 
values. Reserving 10\% of the remaining data for validation further reduces the number of records available for training 
a reliable XGBoost model. Therefore, the XGBoost predictions are discarded, and missing values are imputed using the mean 
of the existing data. Since this approach is not robust, the imputed values are assigned a \textit{LOW} prediction quality.

\textbf{Evaluation of missing value prediction across countries.} As previously mentioned, missing values are observed only in the country-level datasets for Germany or Spain. Consequently, the validation using the 10\% of data set aside specifically assesses how well missing values can be imputed within regions of the same country. Here, the trained models are applied to predict data for regions in the other country, and the results are analyzed.

\begin{figure}[h]
    \centering
    \includegraphics[width=\linewidth]{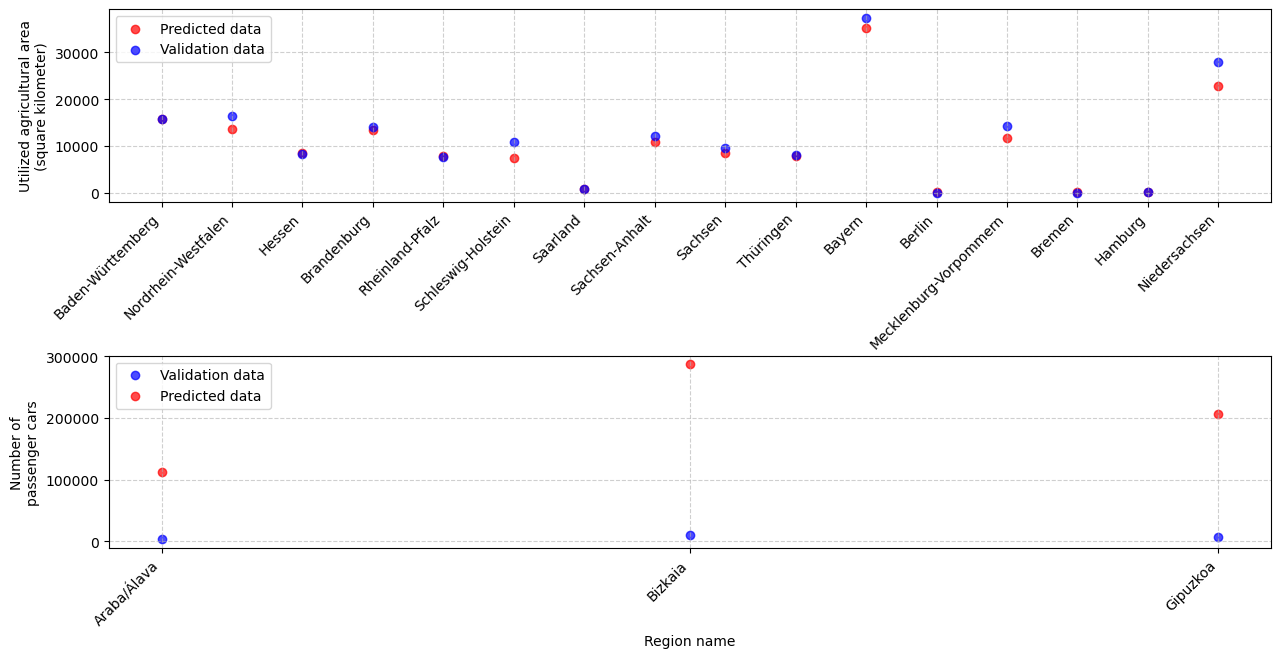}
    \caption{[Top] Results of training an XGBoost model to predict utilized agricultural area in Spain at the LAU level, and applying this model to     estimate values for German LAU regions. The predicted data is compared with the available utilized agricultural area data at the NUTS1 level in Germany. The results indicate that the model's predictions closely align with the actual data, with minimum and maximum deviations of 9.34 and 5106.56 square kilometers, respectively. [Bottom] Results of training an XGBoost model to predict passenger car stock in Germany at the NUTS3 level, and applying this model to 
    estimate values for Spanish NUTS3 regions. The predicted data is compared with the available data for the 3 NUTS3 regions in the 
    Basque Country, Spain. The results indicate that the model's predictions deviate significantly from the actual data, with minimum and maximum 
    deviations of 108957.0 and 276023.0 cars, respectively.}
    \label{fig:uaa_de_cars_es_validation}
\end{figure}

For instance, "utilized agricultural area" is available at the LAU level for Spain. A trained model, initially developed to impute missing values, is also applied to predict values for German LAU regions. These predictions are then validated against data from the Federal Statistical Office of Germany \cite{destatis}, which provides agricultural area figures only at the NUTS1 level. To enable comparison, the predicted values are aggregated accordingly. Figure \ref{fig:uaa_de_cars_es_validation} demonstrates a strong alignment between the predictions and the validation data. This suggests that agricultural land use patterns in Spain and Germany are similar. The distribution of utilized agricultural area in both countries is well explained by highly correlated predictors such as total available area and non-irrigated arable land cover (Figure \ref{fig:lau_es_corr}). Given this successful validation, the XGBoost model is used to impute missing data 
for German LAU regions. 

A similar validation is conducted at the NUTS3 level, where the number of passenger cars per emission group is estimated for 
the Spanish NUTS3 regions. The data is then aggregated into a single dataset representing the total number of passenger cars per NUTS3 region. A further aggregation to the NUTS2 level is performed, for comparison with data from Eustat \cite{eustat_2025}, which provides the number of cars for the three provinces of the Basque Country. Figure \ref{fig:uaa_de_cars_es_validation} presents this comparison, revealing significant deviations between the predictions and the validation data. A similar pattern is observed in other datasets at the German NUTS3 level. These discrepancies suggest that cross-country imputation is not universally reliable, potentially depending on the spatial level of analysis. The availability of more data at the LAU level provides greater variance, allowing for improved learning, whereas similar attempts at the NUTS3 level yield less accurate results. Additionally, sector-specific factors influence the effectiveness of imputation. For example, agricultural indicators exhibit similar spatial distributions in both countries, making cross-country imputation more 
feasible, whereas transport-related indicators do not follow the same pattern. Due to these inconsistencies, the predictions are discarded. 

\subsection*{Stepwise Spatial Disaggregation}

Among the various spatial disaggregation approaches found in the literature, proxy data-based and machine learning-based methods are the most suitable for disaggregating emissions and FEC data \cite{PATIL2024100386}. The proxy data-based approach distributes the target data based on the proportion of the chosen spatial proxy.  In contrast, the machine learning-based approach trains a predictive model, such as XGBoost, to learn the relationships between all available proxy data and the target data at the source spatial level (e.g., NUTS0), and then uses this model to predict the target values in each target region.

Initially, a machine learning-based approach for disaggregation was considered. The approach was eventually discarded due to the 
following reasons:
\begin{enumerate}
    \item The imputation of missing values resulted in poor predictions in certain cases. Applying an additional layer of 
    prediction on top of this may further degrade the results.
    \item In Spain, there are only 52 NUTS3 regions, which may constitute a sample size too small to generate reliable 
    predictions at the LAU level.
    \item Some variable pairs, such as population and gross domestic product, exhibited strong correlations at the NUTS3 
    level but weaker correlations at the LAU level. These differences in correlation raise concerns about whether the 
    statistical relationships among variables, upon which the predictions are based, remain valid at the LAU level.
    \item For most variables, no validation data is available at the LAU level, making it challenging to assess the 
    performance of this disaggregation approach.
\end{enumerate}

Therefore, in this study, a proxy data-based spatial disaggregation method is employed. Here, the quality of disaggregated 
data primarily depends on how effectively the chosen proxy captures the spatial distribution 
of the target data. The selection of a spatial proxy is inherently constrained by the availability of data at fine-scale 
resolution. For each target dataset, potential proxies were initially identified based on theoretical considerations. If the 
most suitable proxy was unavailable in open databases with sufficient non-missing values, the closest alternative was selected. For example, in disaggregating employment data for textile and leather manufacturing, the ideal proxy would be the total size of textile and leather manufacturing facilities in each LAU region. If that data was unavailable, the next best option would be the number of such facilities, followed by a broader proxy such as "industrial or commercial units cover." Since no data on textile and leather manufacturing facilities was accessible, "industrial or commercial units cover" was ultimately chosen as the proxy.

To provide transparency regarding the reliability of the disaggregated data, each proxy is assigned a confidence level \textemdash classified as \textit{HIGH}, \textit{MEDIUM}, \textit{LOW}, or \textit{VERY LOW} \textemdash to indicate its relevance and explanatory strength with respect to the target data. The confidence level reflects the degree of 
alignment between the proxy and the target dataset. For instance, in the example above, the total size of textile and leather 
manufacturing facilities would receive a \textit{HIGH} confidence rating, the number of such facilities would receive a \textit{MEDIUM} rating, and "industrial or commercial units cover" would be rated as \textit{LOW}.

The confidence level assigned to the final disaggregated values at the LAU level is determined by taking the minimum 
of the confidence level of the proxy data and that of the proxy assignment. For instance, if a proxy value in a LAU region is of \textit{MEDIUM} confidence, influenced by the missing value imputation, and the proxy assignment is of \textit{LOW} confidence, then the disaggregated value will be assigned a \textit{LOW} confidence. The selection of proxies 
for the stepwise spatial disaggregation process is outlined in the following.

\textbf{1. NUTS3 variables to LAU.} Tables \ref{tab:potential_proxies_common_nuts3_vars}, \ref{tab:potential_proxies_de_nuts3_vars}, and \ref{tab:potential_proxies_es_nuts3_vars} present the NUTS3 variables alongside their potential proxies. The final proxy selection, determined by data availability, is highlighted with color. It is important to note that some proxies are added although they have different measurement units. For example "construction sites cover" and "road network" are expressed in square kilometer and kilometer. To ensure comparability, all variables are first normalized by their maximum values, preserving true zeros while scaling all other values relative to the highest observed value. This normalization allows proxies to be summed without introducing inconsistencies.

The "industrial or commercial units cover" data is sourced from the Corine Land Cover database, which includes only industrial and commercial units spanning 25 hectares or more. Due to this threshold, many regions have zero values. Since this limitation is consistent across all regions, data imputation was not feasible. Consequently, this proxy is used in conjunction with population data in this study. Furthermore, "employment in construction" uses "construction sites cover" and "road network" as proxies because construction encompasses both building and road construction. 

\textbf{2. NUTS2 variables to LAU.} Table \ref{tab:potential_proxies_common_nuts2_vars} details the proxy assignment process in the case of the NUTS2 variables "number of motorcycles", "air transport of passengers", and "air transport of freight". 

\textbf{3b. FEC (NUTS0 data) to LAU.} Table \ref{tab:potential_proxies_common_fec} presents the FEC end-use sectors for which final proxies were available in both countries. Table \ref{tab:potential_proxies_de_fec} lists the FEC end-use sectors for which final proxies were available exclusively for Germany. Here, emissions from passenger car road transport are disaggregated according to the passenger car fleet categorized into different emission groups. These groups define the vehicle emission standards used in Europe \cite{wikipedia:european_emission_standards}, with each group setting caps on specific air pollutants. The initial emission group, Euro 1, was introduced in July 1992. Over the years, the standards have become increasingly stringent with the introduction of new emission caps. The caps for diesel passenger cars concerning pollutants such as carbon monoxide ($CO$), hydrocarbons and nitrogen oxides ($HC + NO_{X}$), and particulate matter ($PM$) are detailed in Table \ref{tab:emission_standards}. For each emission group, the caps for these three pollutants are summed to obtain a weighting factor for the proxies. The passenger car data provides information for emission group 5 but does not differentiate between Euro 5a and 5b. In this case, the more lenient tier, Euro 5a, is considered to assign more emissions to cars in tier 5. The data also includes an emission group labeled "Other." Due to the lack of additional information from the data source regarding this category, it is treated as the Euro 1 group. Since this data was unavailable for Spain, "average daily traffic - light duty vehicles" is used as a proxy. Similarly, Table \ref{tab:potential_proxies_es_fec} outlines the FEC end-use sectors for which final proxies were available for Spain.

\textbf{3a. GHG Emissions (NUTS0 data) to LAU.} Tables \ref{tab:potential_proxies_common_ghg}, \ref{tab:potential_proxies_de_ghg}, and \ref{tab:potential_proxies_es_ghg} present the proxy assignments in the case of emissions end-use sectors. The proxies are similar to those used for FEC. The differences arise from a different breakdown of the source sub-sectors.

In Germany, except for energy-intensive industries, all proxies correspond directly to relevant emission sources. As a result, most emissions end-use sector proxies are classified as having \textit{HIGH} confidence (see Table \ref{tab:potential_proxies_de_ghg}). In contrast, Spain lacks detailed employment data and spatial data on residential and non-residential areas, limiting the availability of \textit{HIGH} confidence proxies for several emissions end-use sectors (see Table \ref{tab:potential_proxies_es_ghg}).

\section*{Data Records}

The final energy consumption and emissions data at the LAU level for each sub-sector in Germany and Spain are accessible 
on \href{https://zenodo.org/records/14097217?token=eyJhbGciOiJIUzUxMiJ9.eyJpZCI6Ijk0MGJhMGE4LTBmOGUtNDFmYi04M2UzLTZlNGZhMGY5Mjk2MiIsImRhdGEiOnt9LCJyYW5kb20iOiJkYTVkMjAwZmQzYWE5MWY0YzEzM2IzMmFlYmQyNGZhOCJ9.tRU25Dju8BIJXrcmoWsNmmmO6nyVmNo_dsQz0wCjR9KHommVrDOVwLdO9zGGcD1L4klNdqHcBGEThh4CMfyODA}{Zenodo} as .csv files. 
This repository also includes a readme file detailing the repository structure, column definitions, measurement units, and other relevant information.

The spatial disaggregation workflow is being expanded to encompass all 27 EU member states, utilizing \href{https://snakemake.readthedocs.io/en/stable/}{Snakemake} as a workflow manager. Data is regularly updated through the
\href{http://data.localised-project.eu/dsp/docs/}{LOCALISED Data Sharing Platform API}. The spatial proxies introducted in this work are also accessible through this platform, both at their original resolution and in their stepwise disaggregated form down to the LAU level. These disaggregated proxies may be of particular interest to the research data community.

\section*{Technical Validation}

This study introduces a spatial disaggregation workflow that requires technical validation at two critical stages: 
(1) the imputation of missing values in proxy data, and (2) the final disaggregation of FEC and emissions data. 
The validation of the missing value imputation was shown already in the Methods section. 

Validating the disaggregated data is more challenging due to the absence of data on energy consumption and emissions at municipal level in official databases. This lack of data is the primary reason for performing spatial disaggregation of national data. 
Here, technical validation is carried out through the following approaches:
\begin{enumerate}
    \item \textbf{City-level inventories:} Bottom-up inventories reported by selected Spanish cities are used for validation.
    \item \textbf{Cross-validation with disaggregated product}: The results are evaluated against another spatially disaggregated dataset, namely EDGAR. While such datasets are useful for comparative analysis, none provide comprehensive coverage of all the sub-sectors addressed in this study at the municipal level. Furthermore, these alternative datasets are themselves the outcome of spatial disaggregation processes rather than being grounded in official statistics, and therefore warrant independent critical assessment.
    \item \textbf{Visual assessment:} For sectors lacking direct reference datasets, visual inspections are performed to confirm that the spatial distribution of emissions aligns with the patterns of the proxy data used.
\end{enumerate}

The results are discussed in the following subsections.

\textbf{City-level inventories.} We compare the disaggregated results with the FEC and emissions reported by seven Spanish cities \textemdash Barcelona, Madrid, Valencia, Valladolid, Vitoria-Gasteiz, Zaragoza, and Seville \textemdash as part of the Climate-Neutral and Smart Cities initiative \cite{noauthor_climate-neutral_2025}. The climate action plans developed by these cities align in terms of baseline year (2019) and sectoral coverage \cite{noauthor_netzerocities_nodate}. Accordingly, we used 2019 national values for disaggregation to ensure comparability with the reported bottom-up inventories for matching end-use sectors. Two sectors \textemdash buildings and road transport \textemdash could be aligned across datasets, and comparisons were therefore limited to these sectors.

Table \ref{tab:TV_buildings_netzero} presents a comparison of the reported and disaggregated values for the building sector, which includes both household and commerce sectors.  In most cases, the absolute deviation in FEC values remains below 20\%, with notable exceptions in Zaragoza and Seville.

In the case of Zaragoza, further investigation revealed that the reported FEC corresponds to the provincial (NUTS3) level rather than the municipal level. This is confirmed by the municipality's SECAP \cite{zaragoza_secap}, where the building sector FEC is reported as 3,664,235 MWh. Our disaggregated estimate for Zaragoza municipality is 3,670,931 MWh \textemdash resulting in a deviation of only 0.18\%. When disaggregated values are summed to represent the entire province, the resulting FEC is 6,187,351 MWh, which deviates by just 7.26\% from the value reported in Table \ref{tab:TV_buildings_netzero}.

In Seville, the discrepancy arises because only residential electricity consumption is reported under the building sector, excluding other significant components such as commercial and non-electrical energy consumption. This leads to a larger deviation from the disaggregated estimate. These findings highlight the value of top-down disaggregation approaches as a complementary tool to bottom-up inventories, particularly in identifying inconsistencies or omissions in local reporting.

Table \ref{tab:TV_buildings_netzero} also presents a comparison of emission values. Although the same proxies were used for disaggregating both FEC and emissions, the disaggregated emission figures show greater deviation from those reported in bottom-up inventories compared to FEC values. This discrepancy can be attributed to differences in the energy mix between national and regional levels. For instance, according to Eurostat's national energy balance data, the commerce sector uses 19.16\% natural gas. In contrast, the share of natural gas usage in the commerce sector in the provinces of Araba, Bizkaia, and Gipuzkoa is 26.32\%, 24.38\%, and 19.71\%, respectively \cite{basque_energy_agency}. These variations in energy mix can significantly impact emission estimates. Therefore, when developing local inventories using top-down datasets, it is crucial to recalculate emissions if the regional energy mix diverges notably from the national average.

The local inventories also include data for the road transport sector; however, the reporting practices for this sector are often unclear and inconsistent. For instance, in Vitoria-Gasteiz, the reported FEC appears to cover only road transport, whereas in Barcelona, railway transport is also included. In contrast, Valencia’s plan explicitly states that only road transport within the city limits is considered. These inconsistencies in sectoral definitions and reporting scope likely contribute to the significant deviations observed in Table \ref{tab:TV_transport_netzero}.

\textbf{Cross-validation with EDGAR.} EDGAR provides disaggregated emissions data by sector at the NUTS2 level for the year 2022. As a first step, we conduct a sectoral comparison to identify categories that align between datasets. Table \ref{tab:TV_EDGAR_sectoral_comparison} lists the sectors identified as comparable, along with the corresponding national totals reported by both EDGAR and Eurostat.

With the exception of the transport sector, all categories exhibit absolute deviations exceeding 20\%. These discrepancies may arise from differences in the inclusion or exclusion of certain sub-sectors. For instance, chemical industry emissions are not reported for Germany in Eurostat, and thus are not considered here. Moreover, emissions from power plants appear to be included under the industrial sector in EDGAR, whereas they are excluded in our categorisation.

\begin{figure}[h]
    \centering
    \includegraphics[width=\linewidth]{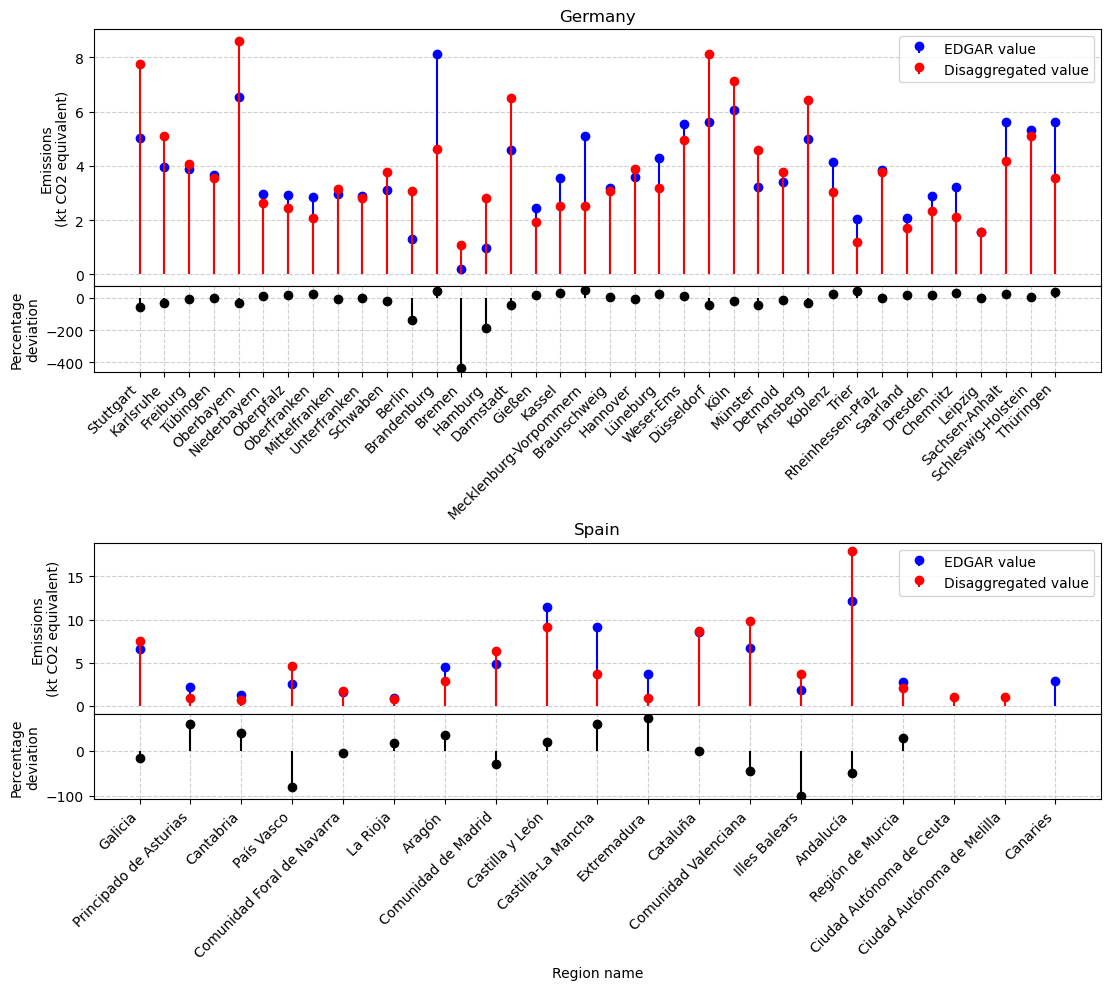}
    \caption{Comparison of transport emissions data available at NUTS2 level from the EDGAR database and the disaggregated values, for Germany and Spain.}
    \label{fig:edgar_transport}
\end{figure}

Given the relatively minor deviation observed in the transport sector at the national scale, a more granular comparison at the NUTS2 level was conducted. To facilitate this, emissions data from all LAU regions were aggregated to their corresponding NUTS2 regions. Figure \ref{fig:edgar_transport} presents this comparison, alongside the associated percentage deviations.

Despite the low national-level discrepancies, significant regional variations are evident across both countries in the two datasets. These regional discrepancies can be attributed to several factors:

\begin{enumerate}
    \item \textit{Propagation of national-level differences:} Although national totals exhibit only minor deviations, these discrepancies are distributed across regions, potentially amplifying inconsistencies at the sub-national level.
    
    \item \textit{Differences in regional coverage:}  In the case of Spain, the Canary Islands were excluded from our analysis due to the unavailability of suitable proxy datasets. Conversely, EDGAR includes this NUTS2 region but omits the autonomous cities of Ceuta and Melilla, which are accounted for in our dataset.
    
    \item \textit{Choice and availability of spatial proxies:} Our approach leverages openly available regional-level datasets as proxies for spatial disaggregation. While EDGAR also employs open data sources, the specific proxies used are not always transparent, making it difficult to isolate the precise causes of spatial discrepancies between the datasets.
\end{enumerate}

\begin{figure}[h]
    \centering
    \includegraphics[width=\linewidth]{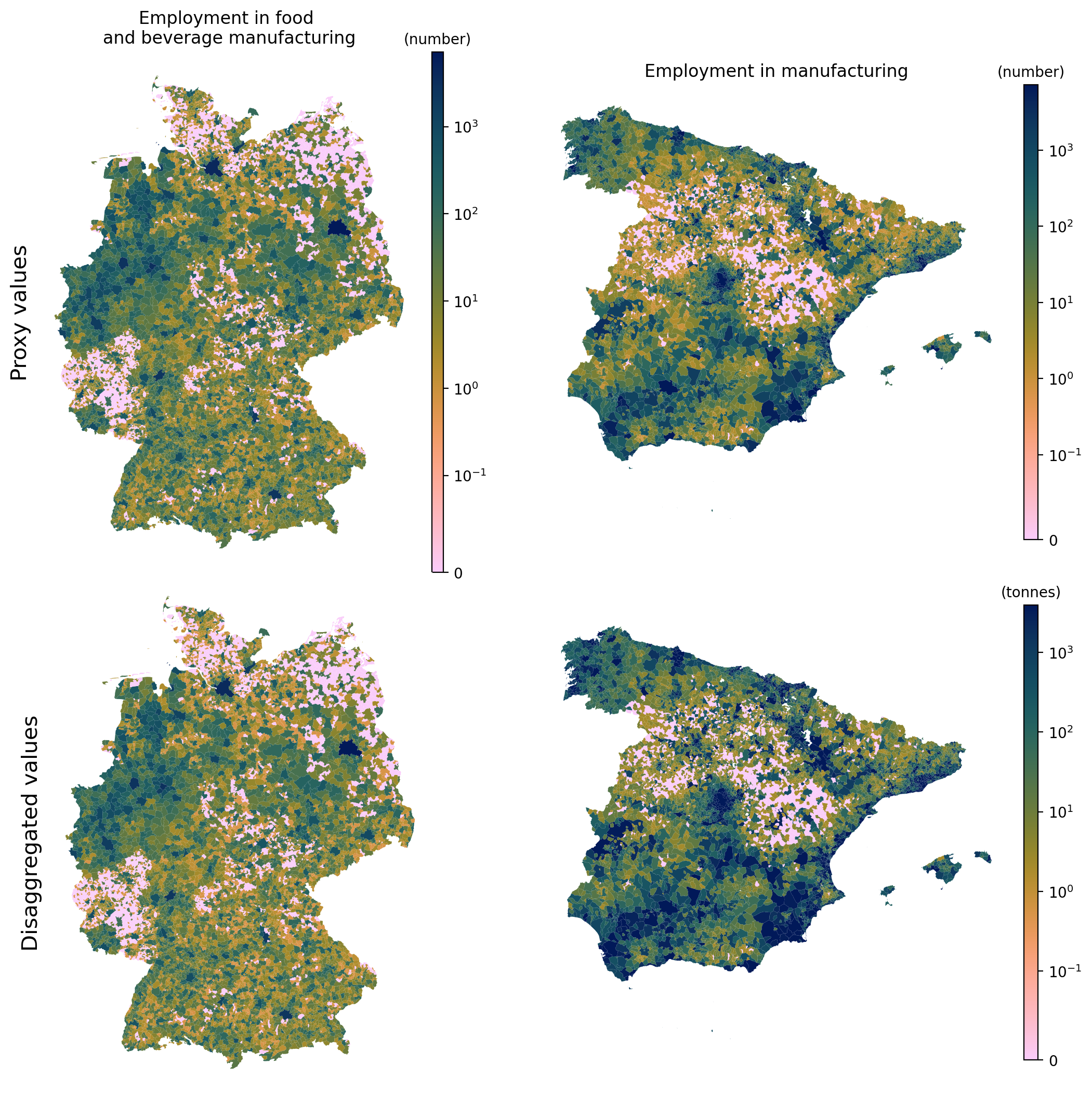}
    \caption{[top-left] "Employment in food and beverage manufacturing" for Germany [top-right] 
    "Employment in manufacturing" in Spain. These proxies are used to 
    disaggregate the emissions in food, beverages, and tobacco industries. [bottom] Disaggregated emission values.}
    \label{fig:non_matching_subsector}
\end{figure}

\textbf{Visual assessment for non-matching sub-sectors.} Figure \ref{fig:non_matching_subsector} illustrates the spatial distribution of the proxy data, namely "employment in food and beverage manufacturing" and "employment in manufacturing" for Germany 
and Spain, respectively. These proxies are used to disaggregate emissions in the food, beverages, and tobacco industries. The figure also displays the disaggregated emission values. It can be observed that the spatial distribution of the disaggregated values mirrors that of the proxy data, confirming that the disaggregation has been performed accurately. Figures pertaining to other sub-sectors are available on GitHub along with the code. The link can be found under the Code Availability section. 

\section*{Usage Notes}
The quality of spatial disaggregation is inherently constrained by the availability and reliability of suitable spatial proxies, particularly with respect to the accuracy of reported values and the extent of missing data. To more precisely evaluate the accuracy of the disaggregation, further comparisons with bottom-up inventories are necessary. However, such inventories are currently limited in number and often exhibit internal inconsistencies. As more consistent and comprehensive bottom-up inventory data becomes available, additional technical validation will be conducted. Corresponding updates to the spatial proxies will also be implemented. All modifications will be documented in the codebase and reflected on the Data Sharing Platform. Further details regarding these resources are provided in the Code Availability section.

For this analysis, the 2019 LAU definitions were applied, acknowledging that LAU boundaries may change annually. Consequently, 
if updated LAU definitions are used in future analyses, data disaggregation may need to be reconfigured for alignment with 
these new regions. Additionally, this adaptation would necessitate reprocessing of any spatial proxy data collected at the 
LAU level. At higher spatial levels, such as NUTS3, boundary definitions typically update on a four-year cycle; for this work, 
we utilized the 2016 NUTS definitions.

The reference year for regional definitions is essential, as is the year of data records, both of which can influence 
disaggregation outcomes. Here, emission and FEC data were collected for 2022. Proxy data, however, come from multiple years, 
with a priority on using the most recent data available from each source. For instance, population data from 2019 aligns 
with the 2019 LAU definitions, while the Corine Land Cover dataset, last updated in 2018, provides land cover information. 
Future updates to this work will include re-running the workflow with the latest available data to ensure accuracy and relevance.

Furthermore, Eurostat offers emissions and Final Energy Consumption (FEC) data for all sub-sectors in both Germany and Spain, with the 
exception of FEC data for the chemical industries in Germany. This specific data has not been available for any of the years
examined. Consequently, we have not disaggregated this value in our work. We will continue to monitor Eurostat for any future updates.

\section*{Code Availability}

The spatial disaggregation workflow developed for this work is implemented in Python and is available on GitHub under the repository 
\href{https://github.com/FZJ-IEK3-VSA/EnergyEmissionsRegio}{EnergyEmissionsRegio}. The core functions can be accessed in the
"energyemissionsregio" directory, while the sections on missing value imputation and disaggregation are in the "experiments" directory. 
This workflow is being expanded to include all 27 EU member states and is regularly updated in the 
\href{https://github.com/FZJ-IEK3-VSA/ETHOS.zoomin}{ETHOS.zoomin} repository on GitHub.

The disaggregation process leverages spatial proxies, which are collected, processed, and stored in a database, where the disaggregated 
data is also saved. This data can be accessed through the LOCALISED Data Sharing Platform API. Additionally, a Python API client for 
accessing this data, named \href{https://github.com/FZJ-IEK3-VSA/LOCALISED-Datasharing-API-Client}{LOCALISED-Datasharing-API-Client}, is available on GitHub.

\bibliography{main}

\section*{Acknowledgements} 

This work was developed as part of the project LOCALISED —Localised decarbonization pathways for citizens, local administrations and
businesses to inform for mitigation and adaptation action. This project received funding from the European Union's Horizon 2020 research
and innovation programme under grant agreement No. 101036458. We extend our sincere gratitude to the entire LOCALISED project team
for their invaluable contributions. We also wish to acknowledge the financial support that has made this project possible. 
This work was also supported by the Helmholtz Association as part of the program “Energy System Design”. Special thanks
are due to our colleagues at the Forschungszentrum Jülich for their diligent proofreading and insightful feedback, which have significantly
enhanced the quality of our work.

\textit{Disclaimer}: This work reflects the authors’ views. The European Commission is not responsible for any use that may be made of the information it
contains.

\section*{Author contributions statement}

S.P. conducted the experiments, analysed the results, and wrote the manuscript, N.P conceived the experiment and supervised the work, J.M.W. provided the resources 
to conduct the work and supervised the work, J.K. supervised the work and was involved in funding acquisition, and D.S. provided the resources to conduct the work. 
All authors reviewed the manuscript. 

\section*{Competing interests}

The authors declare no competing interests.

\section*{Figures \& Tables}

\begin{table}[ht]
    \centering
    \begin{tabular}{|l|l|l|}
    \hline
    \multicolumn{1}{|c|}{\textbf{Data source}} & \multicolumn{1}{c|}{\textbf{Variable}} & \multicolumn{1}{c|}{\textbf{Unit}} \\
    \hline
    Eurostat \cite{lau_statistics_eurostat} & Population & number\\
    \cline{2-3}  
    &  Area & square kilometer \\
    \hline
    Hotmaps \cite{hotmaps_industrial_data} & Number of iron and steel industries & number \\
    \cline{2-3}
     & Number of cement industries & number \\
    \cline{2-3}
    & Number of refineries & number \\
    \cline{2-3}
    & Number of paper and printing industries & number \\
    \cline{2-3}
    & Number of chemical industries & number \\
    \cline{2-3}
    & Number of glass industries & number \\
    \cline{2-3}
    & Number of non-ferrous metals industries & number \\
    \cline{2-3}
    & Number of non-metallic minerals industries & number \\
    \hline
    European Environmental Agency \cite{european_environmental_agency}  &  Average air pollution due to PM2.5 & ug/m3  \\
    \cline{2-3}  
    &  Average air pollution due to NO2 & ug/m3 \\
    \cline{2-3}  
    &  Average air pollution due to O3 & ug/m3 \\
    \cline{2-3}  
    &  Average air pollution due to PM10 & ug/m3 \\
    \hline
    \textcolor{blue}{The National Statistics Institute of Spain} \cite{ine_2024}  & \textcolor{blue}{Utilized agricultural area} & square kilometer \\
    \hline
    \end{tabular}
    \caption{\label{tab:lau_proxy_data_1}LAU-level proxy data collected from Eurostat, Hotmaps, and The National Statistics 
    Institute of Spain. The variables highlighted in blue are available only for Spain.}
\end{table}

\begin{table}[ht]
    \centering
    \begin{tabular}{|l|l|l|}
    \hline
    \multicolumn{1}{|c|}{\textbf{Data source}} & \multicolumn{1}{c|}{\textbf{Variable}} & \multicolumn{1}{c|}{\textbf{Unit}} \\
    \hline
    Corine Land Cover \cite{corine_land_cover} & Continuous urban fabric cover & square kilometer\\
    \cline{2-3}
    & Discontinuous urban fabric cover  & square kilometer \\
    \cline{2-3}
    &  Industrial or commercial units cover  & square kilometer \\
    \cline{2-3}
    &  Port areas cover  & square kilometer  \\
    \cline{2-3}
    &  Airports cover  & square kilometer  \\
    \cline{2-3}
    &  Mineral extraction sites cover  & square kilometer  \\
    \cline{2-3}
    &  Dump sites cover  & square kilometer  \\
    \cline{2-3}
    &  Construction sites cover  & square kilometer  \\
    \cline{2-3}
    &  Green urban areas cover  & square kilometer  \\
    \cline{2-3}
    &  Sport and leisure facilities cover  & square kilometer  \\
    \cline{2-3}
    &  Non irrigated arable land cover  & square kilometer  \\
    \cline{2-3}
    &  Permanently irrigated land cover  & square kilometer  \\
    \cline{2-3}
    &  Rice fields cover  & square kilometer  \\
    \cline{2-3}
    &  Vineyards cover  & square kilometer  \\
    \cline{2-3}
    &  Fruit trees and berry plantations cover  & square kilometer  \\
    \cline{2-3}
    &  Olive groves cover  & square kilometer  \\
    \cline{2-3}
    &  Pastures cover  & square kilometer  \\
    \cline{2-3}
    &  Permanent crops cover   & square kilometer  \\
    \cline{2-3}
    &  Complex cultivation patterns cover  & square kilometer  \\
    \cline{2-3}
    &  Agriculture with natural vegetation cover  & square kilometer  \\
    \cline{2-3}
    &  Agroforestry areas cover  & square kilometer  \\
    \cline{2-3}
    &  Broad leaved forest cover  & square kilometer  \\
    \cline{2-3}
    &  Coniferous forest cover  & square kilometer  \\
    \cline{2-3}
    &  Mixed forest cover  & square kilometer  \\
    \cline{2-3}
    &  Natural grasslands cover  & square kilometer  \\
    \cline{2-3}
    &  Moors and heathland cover  & square kilometer  \\
    \cline{2-3}
    &  Sclerophyllous vegetation cover  & square kilometer  \\
    \cline{2-3}
    &  Transitional woodland shrub cover  & square kilometer  \\
    \cline{2-3}
    &  Beaches dunes and sand cover  & square kilometer  \\
    \cline{2-3}
    &  Bare rocks cover  & square kilometer  \\
    \cline{2-3}
    &  Sparsely vegetated areas cover  & square kilometer  \\
    \cline{2-3}
    & Burnt areas cover  & square kilometer  \\
    \cline{2-3}
    & Glaciers and perpetual snow cover  & square kilometer  \\
    \cline{2-3}
    & Inland marshes cover  & square kilometer  \\
    \cline{2-3}
    & Peat bogs cover  & square kilometer  \\
    \cline{2-3}
    & Salt marshes cover  & square kilometer  \\
    \cline{2-3}
    & Salines cover  & square kilometer  \\
    \cline{2-3}
    & Intertidal flats cover  & square kilometer  \\
    \cline{2-3}
    & Water courses cover  & square kilometer  \\
    \cline{2-3}
    & Water bodies cover  & square kilometer  \\
    \cline{2-3}
    & Coastal lagoons cover  & square kilometer  \\
    \cline{2-3}
    & Estuaries cover  & square kilometer  \\
    \cline{2-3}
    & Sea and ocean cover  & square kilometer  \\
    \hline
    Eurogeographics \cite{euro_regional_map} & Railway network & kilometer \\
    \hline
    OpenStreetMap \cite{open_street_map} &  Road network & kilometer \\
   \cline{2-3}
    &  Number of buildings  & number \\
    \hline
    \end{tabular}
    \caption{\label{tab:lau_proxy_data_2}LAU-level proxy data collected from Corine Land Cover, Eurogeographics, and OpenStreetMap.}
    \end{table}

\begin{table}[ht]
    \centering
    \begin{tabular}{|l|l|l|l|l|}
    \hline
    \textbf{Industry} & \multicolumn{2}{c|}{\textbf{Hotmaps}} &  \multicolumn{2}{c|}{\textbf{sEEnergies}} \\
    \hline
       & \textbf{Germany} & \textbf{Spain} & \textbf{Germany} & \textbf{Spain} \\
    \hline
    Chemical industries & 118 & 25 & 44 & 8\\
    \hline
    Non-ferrous metals industries & 40 & 4 & 17 & 3\\
    \hline
    Non-metallic minerals industries & 160 & 77 & 289 & 52\\
    \hline
    Paper and printing industries & 140 & 159 & 65 & 59\\
    \hline
    Refineries & 26 & 19 & 9 & 8\\
    \hline
\end{tabular}
\caption{\label{tab:industrial_sites_comparison}Number of different industries as reported by Hotmaps and sEEnergies open databases.}
\end{table}

\begin{table}[ht]
    \centering
    \begin{tabular}{|l|l|l|}
    \hline
    \multicolumn{1}{|c|}{\textbf{Data source}} & \multicolumn{1}{c|}{\textbf{Variable}} & \multicolumn{1}{c|}{\textbf{Unit}} \\
    \hline   
    Eurostat & Gross domestic product \cite{eurostat_gdp} & million Euros \\
    \cline{2-3}
    & Road transport of freight \cite{eurostat_road_transport_of_freight} & Mt \\
    \cline{2-3}    
    & Employment in manufacturing \cite{eurostat_employment} & number \\
    \cline{2-3}
    & Employment in construction \cite{eurostat_employment} & number \\
    \cline{2-3}
    & Employment in agriculture, forestry and fishing \cite{eurostat_employment} & number \\
    \hline
    ESPON \cite{espon_soil_sealing} & Soil sealing  & square kilometer \\
    \hline
    EURO-CORDEX \cite{changeeuro} & Heating degree days & heating degree days \\
    \cline{2-3}
    & Cooling degree days & cooling degree days \\
    \hline
    Food and Agriculture Organization of the UN \cite{gilbert2018global}  &  Number of buffaloes  & number \\
        \cline{2-3}
        &  Number of cattle  & number  \\
        \cline{2-3}
        &  Number of pigs   & number  \\
        \cline{2-3}
        &  Number of sheeps   & number  \\
        \cline{2-3}
        &  Number of chickens   & number  \\
        \cline{2-3}
        &  Number of goats   & number  \\
    \hline
    \textcolor{darkorange}{Bundesagentur für Arbeit} \cite{de_employment}  & \textcolor{darkorange}{Employment in textile and leather manufacturing} & number \\
    \cline{2-3} 
    &  \textcolor{darkorange}{Employment in food and beverage manufacturing} & number\\
    \cline{2-3}  
    &  \textcolor{darkorange}{Employment in mechanical and automotive engineering} & number \\
    \cline{2-3}  
    &  \textcolor{darkorange}{Employment in mechatronics, energy and electrical} & number \\
    \cline{2-3} 
    &  \textcolor{darkorange}{Employment in wood processing} & number\\
    \hline  
    \textcolor{darkorange}{Destatis} \cite{destatis}  & \textcolor{darkorange}{Number of passenger cars emission group euro 1} & number \\
    \cline{2-3} 
    & \textcolor{darkorange}{Number of passenger cars emission group euro 2} & number \\
    \cline{2-3} 
    & \textcolor{darkorange}{Number of passenger cars emission group euro 3} & number \\
    \cline{2-3} 
    & \textcolor{darkorange}{Number of passenger cars emission group euro 4} & number \\
    \cline{2-3} 
    & \textcolor{darkorange}{Number of passenger cars emission group euro 5} & number\\
    \cline{2-3} 
    & \textcolor{darkorange}{Number of passenger cars emission group euro 6r} & number \\
    \cline{2-3} 
    & \textcolor{darkorange}{Number of passenger cars emission group euro 6dt} & number\\
    \cline{2-3} 
    & \textcolor{darkorange}{Number of passenger cars emission group euro 6d} & number\\
    \cline{2-3} 
        & \textcolor{darkorange}{Number of passenger cars emission group euro other} & number \\
    \cline{2-3} 
    & \textcolor{darkorange}{Residential building living area} & square kilometer \\
    \cline{2-3} 
    & \textcolor{darkorange}{Non-residential building living area} & square kilometer \\
    \hline    
    \textcolor{blue}{The National Statistics Institute of Spain} \cite{ine_2024}  & \textcolor{blue}{Number of commercial and service companies} & number \\
    \hline    
    \textcolor{blue}{DATAESTUR} \cite{dataestur_2025}  & \textcolor{blue}{Average daily traffic - light duty vehicles} & number \\ 
    \hline
    \end{tabular}
    \caption{\label{tab:nuts3_proxy_data}NUTS3-level proxy data collected from different data sources. 
    The variables highlighted in orange and blue are available only for Germany and Spain, respectively.}
    \end{table}

\begin{table}[ht]
    \centering
    \begin{tabular}{|l|l|l|}
    \hline
    \multicolumn{1}{|c|}{\textbf{Data source}} & \multicolumn{1}{c|}{\textbf{Variable}} & \multicolumn{1}{c|}{\textbf{Unit}} \\
    \hline   
    Eurostat & Number of motorcycles \cite{eurostat_vehicle_stock} & number \\
    \cline{2-3}
    &  Air transport of passengers \cite{eurostat_air_freight_transport} & number \\
    \cline{2-3}
    &  Air transport of freight \cite{eurostat_air_passenger_transport} & Mt \\
    \hline
    \end{tabular}
    \caption{\label{tab:nuts2_proxy_data}NUTS2-level proxy data collected from Eurostat.}
    \end{table}

\begin{table}[ht]
    \centering
    \begin{tabular}{|l|l|l|l|}
    \hline
    \multicolumn{1}{|c|}{\textbf{Spatial level}} & \multicolumn{1}{c|}{\textbf{Variable}} & 
                            \multicolumn{2}{c|}{\textbf{Number of missing values}} \\
      &  & \multicolumn{2}{c|}{\textbf{ (Percentage of missing values)}} \\
     \hline
    \textbf{} & \textbf{} & \multicolumn{1}{c|}{\textbf{Germany}} &  \multicolumn{1}{c|}{\textbf{Spain}}\\
    \hline   
    LAU & \textcolor{blue}{Utilized agricultural area} & - & 348 (4.33\%) \\
    \hline
    NUTS3 & \textcolor{blue}{Number of commerical and service companies} & - & 3 (5.77\%) \\
    \cline{2-4}  
        & \textcolor{blue}{Average daily traffic - light duty vehicles} & - & 10 (19.23\%) \\
    \cline{2-4}    
     & \textcolor{darkorange}{Employment in textile and leather manufacturing} & 34 (8.48\%) & -\\
    \cline{2-4}  
    &  \textcolor{darkorange}{Employment in food and beverage manufacturing} & 2 (0.5\%) & -\\
    \cline{2-4} 
    & \textcolor{darkorange}{Employment in mechanical and automotive engineering} & 1 (0.25\%) & -\\
    \cline{2-4}  
    & \textcolor{darkorange}{Employment in mechatronics, energy and electrical} & 1 (0.25\%) & -\\
    \cline{2-4} 
    & \textcolor{darkorange}{Employment in wood processing} & 2 (0.5\%)  & -\\
    \cline{2-4} 
    & \textcolor{darkorange}{Number of passenger cars emission group euro 1} & 2 (0.50\%) & -\\
    \cline{2-4} 
    & \textcolor{darkorange}{Number of passenger cars emission group euro 2} & 2 (0.50\%) & -\\
    \cline{2-4}
    & \textcolor{darkorange}{Number of passenger cars emission group euro 3} & 2 (0.50\%) & -\\
    \cline{2-4}
    & \textcolor{darkorange}{Number of passenger cars emission group euro 4} & 2 (0.50\%) & -\\
    \cline{2-4}
   & \textcolor{darkorange}{Number of passenger cars emission group euro 5}  & 2 (0.50\%) & -\\
    \cline{2-4} 
   & \textcolor{darkorange}{Number of passenger cars emission group euro 6r}  & 2 (0.50\%) & -\\
    \cline{2-4}
   & \textcolor{darkorange}{Number of passenger cars emission group euro 6dt}  & 2 (0.50\%) & -\\
    \cline{2-4}
   &  \textcolor{darkorange}{Number of passenger cars emission group euro 6d}  & 2 (0.50\%) & -\\
    \cline{2-4} 
    &  \textcolor{darkorange}{Number of passenger cars emission group euro other}  & 2 (0.50\%) & -\\
    \cline{2-4} 
    & \textcolor{darkorange}{Residential building living area}  & 1 (0.25\%) & -\\
    \cline{2-4}
   & \textcolor{darkorange}{Non-residential building living area}  & 1 (0.25\%) & -\\
\hline
    \end{tabular}
    \caption{\label{tab:missing_values}Number of missing values per variable with missing values. The variables 
    highlighted in orange and blue are available only for Germany and Spain, respectively. NOTE: The number of 
    data records at LAU-level in Germany and Spain are 11087 and 8043, respectively. The number of data records 
    at NUTS3-level in Germany and Spain are 401 and 52, respectively.}
    \end{table}

\begin{table}[ht]
    \centering
    \begin{NiceTabular}{p{5cm}llllllll}[hvlines]
    \hline
    \multicolumn{1}{|c|}{\textbf{Variable}} & \multicolumn{4}{c|}{\textbf{Correlation threshold $\ge$ 0.1}} & \multicolumn{4}{c|}{\textbf{Correlation threshold $\ge$ 0.5}} \\
    \hline
     & \multicolumn{2}{c|}{\textbf{Training}} & \multicolumn{2}{c|}{\textbf{Validation}} & \multicolumn{2}{c|}{\textbf{Training}} & \multicolumn{2}{c|}{\textbf{Validation}}\\
    \hline
     &  \multicolumn{1}{|c|}{\textbf{RMSE}}  & \multicolumn{1}{c|}{\textbf{R2}} & \multicolumn{1}{c|}{\textbf{RMSE}}  & \multicolumn{1}{c|}{\textbf{R2}} & \multicolumn{1}{c|}{\textbf{RMSE}} & \multicolumn{1}{c|}{\textbf{R2}}  & \multicolumn{1}{c|}{\textbf{RMSE}} & \multicolumn{1}{c|}{\textbf{R2}}\\
    \hline   
    \textcolor{blue}{Utilized agricultural area} & \cellcolor{green!25}20.31 & \cellcolor{green!25}0.87 & \cellcolor{green!25}20.01 & \cellcolor{green!25}0.89 & 22.01 & 0.85 & 23.28 & 0.85 \\
    \hline   
    \textcolor{blue}{Number of commerical and service companies} & \cellcolor{blue!25}49877.05 & \cellcolor{blue!25}0.58 & \cellcolor{blue!25}27353.46 & \cellcolor{blue!25}0.61 & 47208.70 & 0.66 & 30192.07 & 0.53\\
    \hline  
    \textcolor{blue}{Average daily traffic - light duty vehicles} & \cellcolor{darkred!25}689.45 & \cellcolor{darkred!25}0.18 & \cellcolor{darkred!25}719.52 & \cellcolor{darkred!25}-0.45 & 847.75 & -0.89 & 973.89  &  -1.67 \\
    \hline
    \textcolor{darkorange}{Employment in textile and leather manufacturing} & 290.16 & 0.27 & 529.56 & 0.10 & \cellcolor{darkred!25}311.57 & \cellcolor{darkred!25}0.14 & \cellcolor{darkred!25}505.69 & \cellcolor{darkred!25}0.18\\
    \hline
    \textcolor{darkorange}{Employment in food and beverage manufacturing} & \cellcolor{red!25}737.97 & \cellcolor{red!25}0.22 & \cellcolor{red!25}1482.62 & \cellcolor{red!25}0.29 & 720.42 & 0.29 & 1724.07 & 0.04 \\
    \hline
    \textcolor{darkorange}{Employment in mechanical and automotive engineering} & 2083.65 & 0.79 & 1208.13 & 0.91 & \cellcolor{green!25}1764.42 & \cellcolor{green!25}0.83 & \cellcolor{green!25}1110.13 & \cellcolor{green!25}0.92 \\
    \hline
    \textcolor{darkorange}{Employment in mechatronics, energy and electrical} & 1148.96 & 0.84 & 1023.46 & 0.82 & \cellcolor{green!25}1028.12 & \cellcolor{green!25}0.87 & \cellcolor{green!25}830.13 & \cellcolor{green!25}0.88  \\
    \hline
    \textcolor{darkorange}{Employment in wood processing} & \cellcolor{blue!25}427.62 & \cellcolor{blue!25}0.51  &  \cellcolor{blue!25}259.76 & \cellcolor{blue!25}0.51  & 464.37 & 248.52 & 376.86  & 0.02 \\
    \hline
    \textcolor{darkorange}{Number of passenger cars emission group euro 1} & \cellcolor{green!25}378.22 & \cellcolor{green!25}0.84 & \cellcolor{green!25}155.81 & \cellcolor{green!25}0.92 & 355.74 & 0.84 & 167.48 & 0.91 \\
    \hline
    \textcolor{darkorange}{Number of passenger cars emission group euro 2} & 1481.89 & 0.86 & 758.39 & 0.90 & \cellcolor{green!25}1519.40 & \cellcolor{green!25}0.85 & \cellcolor{green!25}752.10 & \cellcolor{green!25}0.91 \\
    \hline
    \textcolor{darkorange}{Number of passenger cars emission group euro 3} & \cellcolor{green!25}1769.34 & \cellcolor{green!25}0.87 & \cellcolor{green!25}781.02  & \cellcolor{green!25}0.93  & 1785.75 & 0.86 & 1033.32 & 0.87 \\
    \hline
    \textcolor{darkorange}{Number of passenger cars emission group euro 4} & \cellcolor{green!25}7027.27 & \cellcolor{green!25}0.88 & \cellcolor{green!25}3294.12 & \cellcolor{green!25}0.92 & 6786.87 & 0.89 & 3687.20 & 0.90\\
    \hline
    \textcolor{darkorange}{Number of passenger cars emission group euro 5} & 6157.90 & 0.91 & 3386.32  & 0.92 & \cellcolor{green!25}5917.91 & \cellcolor{green!25}0.91 & \cellcolor{green!25}3050.37 & \cellcolor{green!25}0.93\\
    \hline
    \textcolor{darkorange}{Number of passenger cars emission group euro 6r} & 6978.70 & 0.90 & 3579.00 &  0.92 & \cellcolor{green!25}6967.55 & \cellcolor{green!25}0.89 & \cellcolor{green!25}3269.41 & \cellcolor{green!25}0.94\\
    \hline
    \textcolor{darkorange}{Number of passenger cars emission group euro 6dt} & 4320.35 & 0.85 & 2555.25 & 0.87 & \cellcolor{green!25}4317.94 & \cellcolor{green!25}0.85 & \cellcolor{green!25}1461.44 & \cellcolor{green!25}0.95\\
    \hline
    \textcolor{darkorange}{Number of passenger cars emission group euro 6d} & 7072.20 & 0.70 & 8294.98 & 0.52 &  \cellcolor{blue!25}7472.18 &  \cellcolor{blue!25}0.59  &  \cellcolor{blue!25}8127.21 &  \cellcolor{blue!25}0.54\\
    \hline    
    \textcolor{darkorange}{Number of passenger cars emission group euro other} & 2113.87 & 0.78 & 1627.34 & 0.78 & \cellcolor{green!25}2164.24 & \cellcolor{green!25}0.76 & \cellcolor{green!25}1289.55 & \cellcolor{green!25}0.86 \\
    \hline
    \textcolor{darkorange}{Residential building living area} & \cellcolor{green!25}3.26 &  \cellcolor{green!25}0.89 & \cellcolor{green!25}1.03 & \cellcolor{green!25}0.96 & 3.37  & 0.88 & 1.78 & 0.89 \\
    \hline  
    \textcolor{darkorange}{Non-residential building living area} & \cellcolor{green!25}0.09 & \cellcolor{green!25}0.83 & \cellcolor{green!25}0.06 & \cellcolor{green!25}0.91 & 0.08 & 0.84 & 0.07 & 0.88 \\
    \hline
\end{NiceTabular}
\caption{\label{tab:missing_value_imputation_results}The RMSE and R-squared errors on training and validation data. The colored cells indicate 
the best performing model between the two: predictors with correlation threshold $\ge$ 0.1 and predictors with correlation threshold $\ge$ 0.5. The 
color of the cell indicates the confidence rating of the imputed values. See Table \ref{tab:imputation_confidence_level} for confidence rating details.}
\end{table}

\begin{table}[ht]
    \centering
   \begin{NiceTabular}{ll}[hvlines]
    \hline
    \textbf{R-squared threshold} & \textbf{Value confidence level} \\
    \hline
    \cellcolor{green!25}$>0.8$ & \cellcolor{green!25}\textit{HIGH} \\
    \hline
    \cellcolor{blue!25} $>0.5$ and $\leq0.8$ & \cellcolor{blue!25}\textit{MEDIUM}  \\
   \hline
   \cellcolor{red!25} $>0.2$ and $\leq0.5$ & \cellcolor{red!25}\textit{LOW}   \\
    \hline
    \cellcolor{darkred!25} $\leq0.2$ & \cellcolor{darkred!25}\textit{VERY LOW}   \\
    \hline
\end{NiceTabular}
\caption{\label{tab:imputation_confidence_level}The threshold for R-sqaured score and the associated confidence level for imputed values.}
\end{table}  

\begin{table}[ht]
    \centering
    \begin{NiceTabular}{p{4cm}p{4cm}p{4cm}p{4cm}}[hvlines]
    \hline
    \textbf{NUTS3 Variable}    &  \multicolumn{3}{|c|}{\textbf{Potential proxies}}  \\
    \hline
                         & \textbf{Most suitable} & \textbf{Alternative proxy 1} & \textbf{Alternative proxy 2}  \\
    \hline
    Employment in manufacturing & \cellcolor{green!25}Industrial or commercial units cover + Population & - & - \\
    \hline
    Employment in construction & Number of construction companies & \cellcolor{blue!25}Construction sites cover + Road network & - \\
    \hline
    Road transport of freight & Number of heavy duty vehicles & Number of fuel stations & \cellcolor{red!25}Road network \\     
    \hline
    Heating degree days & Temperature & \cellcolor{blue!25}(No proxy. Same value for all child regions) & - \\
    \hline
    Number of cattle &  Size of cattle farms & Number of cattle farms & \cellcolor{red!25} Utilized agricultural area \\ 
    \hline
    Number of pigs & Size of pig farms & Number of pig farms & \cellcolor{red!25} Utilized agricultural area \\ 
    \hline
    Number of buffaloes & Size of buffalo farms & Number of buffalo farms & \cellcolor{red!25} Utilized agricultural area \\ 
    \hline
    Employment in agriculture, forestry, and fishing & Utilized agricultural area + 
                                                                    Agroforestry areas cover +  
                                                                    Number of fishers  &  \cellcolor{blue!25}
                                                                                        Utilized agricultural area + 
                                                                                        Agroforestry areas cover +  
                                                                                        Water bodies cover +
                                                                                        Water courses cover
                                                                                    & - \\ 
    \hline
\end{NiceTabular}
\caption{\label{tab:potential_proxies_common_nuts3_vars} The potential proxies for disaggregating each NUTS3 dataset commonly collected for both Germany and 
Spain are presented. The final chosen proxy is highlighted, with color coding indicating the confidence level. Refer to Table \ref{tab:imputation_confidence_level} 
for the confidence level color scheme.}
\end{table}
\begin{table}[ht]
    \centering
    \begin{NiceTabular}{p{4cm}p{4cm}p{4cm}p{4cm}}[hvlines]
    \hline
    \textbf{NUTS3 Variable}    &  \multicolumn{3}{|c|}{\textbf{Potential proxies}}  \\
    \hline
                         & \textbf{Most suitable} & \textbf{Alternative proxy 1} & \textbf{Alternative proxy 2}  \\
    \hline
    \textcolor{darkorange}{Employment in textile and leather manufacturing} & Size of textile and leather manufacturing industries & Number of textile and leather manufacturing industries & \cellcolor{red!25}Industrial or commercial units cover + Population \\ 
    \hline
    \textcolor{darkorange}{Employment in food and beverage manufacturing} & Size of food and beverage manufacturing industries & Number of food and beverage manufacturing industries & \cellcolor{red!25}Industrial or commercial units cover + Population \\ 
    \hline
    \textcolor{darkorange}{Employment in mechanical and automotive engineering} & Size of mechanical and automotive engineering industries & Number of mechanical and automotive engineering industries & \cellcolor{red!25}Industrial or commercial units cover + Population \\ 
    \hline
    \textcolor{darkorange}{Employment in mechatronics, energy and electrical} & Size of mechatronics, energy and electrical industries & Number of mechatronics, energy and electrical industries & \cellcolor{red!25}Industrial or commercial units cover + Population  \\ 
    \hline
    \textcolor{darkorange}{Employment in wood processing} & Size of wood processing industries & Number of wood processing industries & \cellcolor{red!25}Industrial or commercial units cover + Population \\ 
    \hline
    \textcolor{darkorange}{Number of passenger cars emission group euro 1} & Age distribution of passenger cars &  Number of fuel stations & \cellcolor{red!25}Population  \\ 
    \hline
    \textcolor{darkorange}{Number of passenger cars emission group euro 2} & Age distribution of passenger cars &  Number of fuel stations & \cellcolor{red!25} Population  \\ 
    \hline
    \textcolor{darkorange}{Number of passenger cars emission group euro 3} & Age distribution of passenger cars &   Number of fuel stations & \cellcolor{red!25} Population \\ 
    \hline
    \textcolor{darkorange}{Number of passenger cars emission group euro 4} & Age distribution of passenger cars &   Number of fuel stations & \cellcolor{red!25} Population  \\   
    \hline
    \textcolor{darkorange}{Number of passenger cars emission group euro 5} & Age distribution of passenger cars &  Number of fuel stations & \cellcolor{red!25} Population  \\    
    \hline
    \textcolor{darkorange}{Number of passenger cars emission group euro 6r} & Age distribution of passenger cars &  Number of fuel stations & \cellcolor{red!25} Population   \\ 
    \hline
    \textcolor{darkorange}{Number of passenger cars emission group euro 6dt} & Age distribution of passenger cars &  Number of fuel stations & \cellcolor{red!25} Population  \\  
    \hline
    \textcolor{darkorange}{Number of passenger cars emission group euro 6d} & Age distribution of passenger cars &  Number of fuel stations & \cellcolor{red!25} Population  \\ 
    \hline
    \textcolor{darkorange}{Number of passenger cars emission group euro other} & Age distribution of passenger cars &  Number of fuel stations & \cellcolor{red!25} Population \\ 
    \hline
    \textcolor{darkorange}{Residential building living area} & \cellcolor{green!25} Population & - & - \\ 
    \hline
    \textcolor{darkorange}{Non-residential building living area} & \cellcolor{green!25}Industrial or commercial units cover + Population  & - & - \\ 
    \hline
\end{NiceTabular}
\caption{\label{tab:potential_proxies_de_nuts3_vars}The potential proxies for disaggregating each NUTS3 dataset collected only for Germany are presented. 
The final chosen proxy is highlighted, with color coding indicating the confidence level. Refer to Table \ref{tab:imputation_confidence_level} 
for the confidence level color scheme.}
\end{table}

\begin{table}[ht]
    \centering
    \begin{NiceTabular}{p{4cm}p{4cm}p{4cm}p{4cm}}[hvlines]
    \hline
    \textbf{NUTS3 Variable}    &  \multicolumn{3}{|c|}{\textbf{Potential proxies}}  \\
    \hline
                         & \textbf{Most suitable} & \textbf{Alternative proxy 1} & \textbf{Alternative proxy 2}  \\
    \hline
    \textcolor{blue}{Number of commerical and service companies} & \cellcolor{green!25}Industrial or commercial units cover + Population  & - & - \\
    \hline
    \textcolor{blue}{Average daily traffic - light duty vehicles} & Number of light duty vehicles & Number of fuel stations & \cellcolor{red!25} Population \\ 
    \hline
\end{NiceTabular}
\caption{\label{tab:potential_proxies_es_nuts3_vars}The potential proxies for disaggregating each NUTS3 dataset collected only for Spain are presented. 
The final chosen proxy is highlighted, with color coding indicating the confidence level. Refer to Table \ref{tab:imputation_confidence_level} 
for the confidence level color scheme.}
\end{table}

\begin{table}[ht]
    \centering
    \begin{NiceTabular}{p{4cm}p{4cm}p{4cm}p{4cm}}[hvlines]
    \hline
    \textbf{NUTS2 Variable}    &  \multicolumn{3}{|c|}{\textbf{Potential proxies}}  \\
    \hline
                         & \textbf{Most suitable} & \textbf{Alternative proxy 1} & \textbf{Alternative proxy 2}  \\
    \hline
    Number of motorcycles & Number of motorcycle service stations & Number of fuel stations & \cellcolor{red!25}Road network \\
    \hline
    Air transport of passengers & \cellcolor{green!25}Airports cover & - & - \\
    \hline
    Air transport of freight & \cellcolor{green!25}Airports cover & - & - \\
    \hline
\end{NiceTabular}
\caption{\label{tab:potential_proxies_common_nuts2_vars} The potential proxies for disaggregating each NUTS2 dataset commonly collected for both Germany and 
Spain are presented. The final chosen proxy is highlighted, with color coding indicating the confidence level. Refer to Table \ref{tab:imputation_confidence_level} 
for the confidence level color scheme.}
\end{table}

\begin{table}[ht]
    \centering
    \begin{NiceTabular}{p{4cm}p{4cm}p{4cm}}[hvlines]
    \hline
    \textbf{Emission source}    &  \multicolumn{3}{|c|}{\textbf{Potential proxies}}  \\
    \hline
                         & \textbf{Most suitable} & \textbf{Alternative proxy 1}  \\
    \hline
    Iron and steel industries & Capacity of iron and steel industries & \cellcolor{blue!25}Number of iron and steel industries \\
    \hline
    Non-ferrous metals industries & Capacity of non-ferrous metals industries & \cellcolor{blue!25}Number of non-ferrous metals industries \\
    \hline
    Chemical industries & Capacity of chemical industries & \cellcolor{blue!25}Number of chemical industries \\
    \hline
    Non-metallic minerals industries & Capacity of non-metallic minerals industries & \cellcolor{blue!25}Number of non-metallic minerals industries  \\   
    \hline 
    Mining and quarrying & \cellcolor{green!25}Mineral extraction sites cover & - \\ 
    \hline
    Paper, pulp, and printing industries & Capacity of paper, pulp, and printing industries & \cellcolor{blue!25}Number of paper and printing industries \\  
    \hline
    Construction & \cellcolor{green!25}Employment in construction & - \\
    \hline
    Rail transport &  \cellcolor{green!25}Railway network & - \\
    \hline
    Domestic aviation  & \cellcolor{green!25}Air transport of freight + Air transport of passengers  & -\\    
    \hline
    Domestic navigation  & \cellcolor{green!25}Port areas cover & - \\    
    \hline 
    Agriculture and forestry & \cellcolor{green!25}Employment in agriculture, forestry, and fishing  & - \\  
    \hline
\end{NiceTabular}
\caption{\label{tab:potential_proxies_common_fec} FEC end-use sectors with final proxies commonly available for both Germany and Spain. 
The final chosen proxy is highlighted, with color coding indicating the confidence level. 
Refer to Table \ref{tab:imputation_confidence_level} for the confidence level color scheme.}
\end{table}

\begin{table}[ht]
    \centering
    \begin{NiceTabular}{p{4cm}p{9cm}}[hvlines]
    \hline
    \textbf{FEC source}    &  \textbf{Most suitable proxy}  \\
    \hline
    Wood and wood products industries & \cellcolor{green!25}Employment in wood processing \\
    \hline 
    Transport equipment industries & \cellcolor{green!25}Employment in mechanical and automotive engineering   \\
    \hline 
    Machinery industries & \cellcolor{green!25}Employment in mechatronics, energy and electrical  \\
    \hline 
    Food, beverages, and tobacco industries & \cellcolor{green!25}Employment in food and beverage manufacturing \\
    \hline 
    Textile and leather industries & \cellcolor{green!25}Employment in textile and leather manufacturing  \\
    \hline 
    Road transport & \cellcolor{green!25}(Road transport of freight) +
                                        (3.83 * Number of passenger cars emission group euro 1) + 
                                    (1.78 * Number of passenger cars emission group euro 2) +
                                    (1.25 * Number of passenger cars emission group euro 3) +
                                    (0.825 * Number of passenger cars emission group euro 4) +
                                    (0.735 * Number of passenger cars emission group euro 5) +
                                    (0.6745 * Number of passenger cars emission group euro 6r) +
                                    (0.6745 * Number of passenger cars emission group euro 6dt) +
                                    (0.6745 * Number of passenger cars emission group euro 6d) +
                                    (3.83 * Number of passenger cars emission group euro other)  \\
    \hline
    Households & \cellcolor{green!25}Residential building living area * Heating degree days \\
    \hline 
    Commerce & \cellcolor{green!25}Non-residential building living area * Heating degree days  \\
\hline
\end{NiceTabular}
\caption{\label{tab:potential_proxies_de_fec} FEC end-use sectors with final proxies available for 
Germany. The final chosen proxy is highlighted, with color coding indicating the confidence level. 
Refer to Table \ref{tab:imputation_confidence_level} for the confidence level color scheme.}
\end{table}

\begin{table}
    \centering
    \begin{tabular}{|l|l|l|l|l|}
        \hline
        \multicolumn{1}{|c|}{\textbf{Tier}} & \multicolumn{1}{c|}{$\textbf{CO}$}
                            & \multicolumn{1}{c|}{$\textbf{HC+NO}_{\textbf{X}}$} 
                            & \multicolumn{1}{c|}{$\textbf{PM}$} 
                            & \multicolumn{1}{c|}{\textbf{Total}} \\
        \hline
        Euro 1 & 2.72 & 0.97 & 0.14  & 3.83\\
        \hline
        Euro 2 & 1.0 & 0.7 & 0.08 & 1.78\\
        \hline
        Euro 3 & 0.66 & 0.56 & 0.05 & 1.25\\
        \hline
        Euro 4 & 0.50 & 0.30 & 	0.025 & 0.825\\
        \hline
        Euro 5a & 0.50 & 0.230 & 0.005 & 0.735 \\
        \hline
        Euro 5b & 0.50 & 0.230 & 0.0045 & 0.7345\\
        \hline
        Euro 6b & 0.50 & 0.170 & 0.0045 & 0.6745\\
        \hline
        Euro 6c & 0.50 & 0.170 & 0.0045 & 0.6745\\
        \hline
        Euro 6d-temp & 0.50 & 0.170 & 0.0045 & 0.6745\\
        \hline
        Euro 6d & 0.50 & 0.170 & 0.0045 & 0.6745\\
        \hline
        Euro 6e & 0.50 & 0.170 & 0.0045 & 0.6745\\
        \hline
        \end{tabular}
    \caption{\label{tab:emission_standards}Emission caps for different air pollutants, per emission group.}
\end{table}

\begin{table}[ht]
    \centering
    \begin{NiceTabular}{p{4cm}p{4cm}p{4cm}}[hvlines]
    \hline
    \textbf{FEC source}    &  \multicolumn{3}{|c|}{\textbf{Potential proxies}}  \\
    \hline
                         & \textbf{Most suitable} & \textbf{Alternative proxy 1}  \\
    \hline
    Wood and wood products industries & Employment in wood processing & \cellcolor{blue!25}Employment in manufacturing \\
    \hline 
    Transport equipment industries & Employment in mechanical and automotive engineering  & \cellcolor{blue!25}Employment in manufacturing \\
    \hline 
    Machinery industries & Employment in mechatronics, energy and electrical & \cellcolor{blue!25}Employment in manufacturing \\
    \hline 
    Food, beverages, and tobacco industries & Employment in food and beverage manufacturing & \cellcolor{blue!25}Employment in manufacturing\\
    \hline 
    Textile and leather industries & Employment in textile and leather manufacturing & \cellcolor{blue!25}Employment in manufacturing \\ 
    \hline 
    Road transport & \cellcolor{green!25}Road transport of freight + Average daily traffic - light duty vehicles & - \\
    \hline
    Households & Residential building living area * Heating degree days & \cellcolor{blue!25}Population * Heating degree days \\
    \hline 
    Commerce & Non-residential building living area * Heating degree days & \cellcolor{blue!25}Number of commercial and service companies * Heating degree days \\
\hline
\end{NiceTabular}
\caption{\label{tab:potential_proxies_es_fec} FEC end-use sectors with final proxies available for 
Spain. The final chosen proxy is highlighted, with color coding indicating the confidence level. 
Refer to Table \ref{tab:imputation_confidence_level} for the confidence level color scheme.}
\end{table}

\begin{table}[ht]
    \centering
    \begin{NiceTabular}{p{4cm}p{4cm}p{4cm}}[hvlines]
    \hline
    \textbf{Emission source}    &  \multicolumn{3}{|c|}{\textbf{Potential proxies}}  \\
    \hline
                         & \textbf{Most suitable} & \textbf{Alternative proxy 1}  \\
    \hline
    Iron and steel industries & Capacity of iron and steel industries & \cellcolor{blue!25}Number of iron and steel industries \\
    \hline
    Non-ferrous metals industries & Capacity of non-ferrous metals industries & \cellcolor{blue!25}Number of non-ferrous metals industries \\
    \hline
    Chemical industries & Capacity of chemical industries & \cellcolor{blue!25}Number of chemical industries \\
    \hline
    Non-metallic minerals industries & Capacity of non-metallic minerals industries & \cellcolor{blue!25}Number of non-metallic minerals industries  \\   
    \hline 
    Paper, pulp, and printing industries & Capacity of paper, pulp, and printing industries & \cellcolor{blue!25}Number of paper and printing industries \\  
    \hline
    Rail transport &  \cellcolor{green!25}Railway network & - \\
    \hline
    Road freight transport &  \cellcolor{green!25}Road transport of freight & - \\     
    \hline
    Road transport using motorcycles & \cellcolor{green!25}Number of motorcycles  & - \\    
    \hline
    Domestic aviation  & \cellcolor{green!25}Air transport of freight + Air transport of passengers  & -\\    
    \hline
    Domestic navigation  & \cellcolor{green!25}Port areas cover & - \\    
    \hline 
    Cultivation & \cellcolor{green!25}Utilized agricultural area  & - \\  
    \hline
    Livestock  & \cellcolor{green!25}Number of cattle + Number of pigs + Number of buffaloes & -  \\  
\end{NiceTabular}
\caption{\label{tab:potential_proxies_common_ghg} GHG emissions end-use sectors with final proxies commonly available for both Germany and Spain. 
The final chosen proxy is highlighted, with color coding indicating the confidence level. 
Refer to Table \ref{tab:imputation_confidence_level} for the confidence level color scheme.}
\end{table}
\begin{table}[ht]
    \centering
    \begin{NiceTabular}{p{4cm}p{9cm}}[hvlines]
    \hline
    \textbf{Emission source}    &  \textbf{Most suitable proxy}  \\
    \hline
    Food, beverages, and tobacco industries & \cellcolor{green!25}Employment in food and beverage manufacturing \\
    \hline
    Other manufacturing industries and construction & \cellcolor{green!25}Employment in mechanical and automotive engineering + 
                                                                        Employment in mechatronics, energy and electrical 
                                                                        + Employment in textile and leather manufacturing 
                                                                        + Employment in construction  \\
    \hline    
    Road transport using cars   &  \cellcolor{green!25} (3.83 * Number of passenger cars emission group euro 1) + 
                                                        (1.78 * Number of passenger cars emission group euro 2) +
                                                        (1.25 * Number of passenger cars emission group euro 3) +
                                                        (0.825 * Number of passenger cars emission group euro 4) +
                                                        (0.735 * Number of passenger cars emission group euro 5) +
                                                        (0.6745 * Number of passenger cars emission group euro 6r) +
                                                        (0.6745 * Number of passenger cars emission group euro 6dt) +
                                                        (0.6745 * Number of passenger cars emission group euro 6d) +
                                                        (3.83 * Number of passenger cars emission group euro other) \\
    \hline
    Households & \cellcolor{green!25}Residential building living area * Heating degree days \\
    \hline 
    Commerce & \cellcolor{green!25}Non-residential building living area * Heating degree days  \\
    \hline
\end{NiceTabular}
\caption{\label{tab:potential_proxies_de_ghg} GHG emissions end-use sectors with final proxies available for Germany. 
The final chosen proxy is highlighted, with color coding indicating the confidence level. 
Refer to Table \ref{tab:imputation_confidence_level} for the confidence level color scheme.}
\end{table}

\begin{table}[ht]
    \centering
    \begin{NiceTabular}{p{4cm}p{6cm}p{6cm}}[hvlines]
    \hline
    \textbf{Emission source}    &  \multicolumn{3}{|c|}{\textbf{Potential proxies}}  \\
    \hline
             & \textbf{Most suitable} & \textbf{Alternative proxy 1}  \\
    \hline
    Food, beverages, and tobacco industries & Employment in food and beverage manufacturing & \cellcolor{blue!25}Employment in manufacturing\\
    \hline
    Other manufacturing industries and construction & Employment in mechanical and automotive engineering + 
                                                                        Employment in mechatronics, energy and electrical 
                                                                        + Employment in textile and leather manufacturing 
                                                                        + Employment in construction & \cellcolor{blue!25}Employment in manufacturing 
                                                                        + Employment in construction\\
    \hline    
    Road transport using cars   &  \cellcolor{green!25}Average daily traffic - light duty vehicles & - \\
    \hline
    Households & Residential building living area * Heating degree days & \cellcolor{blue!25}Population * Heating degree days \\
    \hline 
    Commerce & Non-residential building living area * Heating degree days & \cellcolor{blue!25}Number of commerical and service companies * Heating degree days \\
\end{NiceTabular}
\caption{\label{tab:potential_proxies_es_ghg} GHG emissions end-use sectors with final proxies available for Spain. 
The final chosen proxy is highlighted, with color coding indicating the confidence level. 
Refer to Table \ref{tab:imputation_confidence_level} for the confidence level color scheme.}
\end{table}

\begin{table}
    \centering
    \begin{tabular}{|p{1.5cm}|p{2.5cm}|p{2.5cm}|p{2.5cm}|p{2.5cm}|p{2.5cm}|}
        \hline
        & \textbf{City} & \textbf{Reported value (MWh)}  & \textbf{Disaggregated value (MWh)} & \textbf{Difference (MWh)} & \textbf{Percentage deviation (\%)} \\
        \hline
        \textbf{FEC} & Barcelona & 9822750 & 8841562 & 981188 & 9.99\\ 
        \cline{2-6}
        & Madrid  & 21644328 & 21892150 & -247822 & -1.14\\ 
        \cline{2-6}
        & Valencia  & 3102478.65 & 3238758 & -136279.35 & -4.39\\ 
        \cline{2-6}
        & Valladolid  & 1703105.56 & 1970336 & -267230.44 & -15.69\\ 
        \cline{2-6}
        & Vitoria-Gasteiz  & 1819780 & 1592324 & 227456 & 12.50\\ 
       \cline{2-6}
        & Zaragoza  & 5768598.63 & 3670931 & 2097667.63 & 36.36\\ 
        \cline{2-6}
        & Seville  & 1078192.20 & 2166460.60 & -1088267.8 & -100.93\\
        \hline
        \hline
        & \textbf{City} & \textbf{Reported value (kt CO2 equivalent)}  & \textbf{Disaggregated value (kt CO2 equivalent)} & \textbf{Difference (kt CO2 equivalent)} & \textbf{Percentage deviation (\%)} \\
        \hline
        \textbf{Emissions} & Barcelona & 813 & 806.18 & 6.82 & 0.84\\ 
        \cline{2-6}
        & Madrid  & 2130.89 & 1975.78 & 155.11 & 7.28\\ 
        \cline{2-6}
        & Valencia  & 303.94 & 298.26 & 5.68 & 1.87\\ 
        \cline{2-6}
        & Valladolid  & 259.25 & 179.05 & 80.2 & 30.94\\ 
        \cline{2-6}
        & Vitoria-Gasteiz  & 230.18 & 145 & 84.73 & 36.81\\ 
       \cline{2-6}
        & Zaragoza  & 900.42 & 333.74 & 566.68 & 62.94 \\ 
        \cline{2-6}
        & Seville  & 127.83 & 198.01 & -70.18 & -54.90\\
        \hline
        \end{tabular}
    \caption{\label{tab:TV_buildings_netzero} Comparison of FEC and emissions values reported by seven Spanish cities with the disaggregated values for the building sector. NOTE: Building sector includes households and commerce sectors.}
\end{table}

\begin{table}
    \centering
    \begin{tabular}{|p{1.5cm}|p{2.5cm}|p{2.5cm}|p{2.5cm}|p{2.5cm}|p{2.5cm}|}
        \hline
        & \textbf{City} & \textbf{Reported value (MWh)}  & \textbf{Disaggregated value (MWh)} & \textbf{Difference (MWh)} & \textbf{Percentage deviation (\%)} \\
        \hline
        \textbf{FEC} & Barcelona & 3416400 & 1108623 & 2307777 & 67.55\\ 
        \cline{2-6}
        & Madrid  & 11160182 & 6636013 & 4524169 & 40.54\\ 
        \cline{2-6}
        & Valencia  & 4567565.19 & 3238758 & 1920795.19 & 42.05\\ 
        \cline{2-6}
        & Valladolid  & 2188477.62 & 2646770 & 537697.62 & 24.57\\ 
        \cline{2-6}
        & Vitoria-Gasteiz  & 1819780 & 2664792 & -1849896 & -227.01\\ 
       \cline{2-6}
        & Zaragoza  & 1859814.25 & 3644826 & -1785011.75 & -95.98\\ 
        \cline{2-6}
        & Seville  & 3330893.6 & 2109571 & 1221322.6 & 36.67\\
        \hline
        \hline
        & \textbf{City} & \textbf{Reported value (kt CO2 equivalent)}  & \textbf{Disaggregated value (kt CO2 equivalent)} & \textbf{Difference (kt CO2 equivalent)} & \textbf{Percentage deviation (\%)} \\
        \hline
        \textbf{Emissions} & Barcelona & 906 & 519,03 & 386.97 & 42.71\\ 
        \cline{2-6}
        & Madrid  & 2315.24 & 2429.12 & -113.88 & 4.92\\ 
        \cline{2-6}
        & Valencia  & 513.58 & 1235.63 & -722.05 & 140.59\\ 
        \cline{2-6}
        & Valladolid  & 231.99 & 808.54 & -576.55 & 248.52\\ 
        \cline{2-6}
        & Vitoria-Gasteiz  & 197.26 & 983.59 & -786.33 & 398.63\\ 
       \cline{2-6}
        & Zaragoza  & 480.67 & 333.74 & -525.34 & 109.29 \\ 
        \cline{2-6}
        & Seville  & 789.59 & 703.81 & 85.78 & 10.86\\
        \hline
        \end{tabular}
    \caption{\label{tab:TV_transport_netzero} Comparison of FEC and emissions values reported by seven Spanish cities with the disaggregated values for the road transport sector.}
\end{table}

\begin{table}[ht]
    \centering
    \begin{NiceTabular}{p{2cm}p{1.5cm}p{1.5cm}p{1.5cm}p{1.5cm}p{1.5cm}p{1.5cm}}[hvlines]
    \hline
    \textbf{Sector}   &  \multicolumn{3}{|c}{\textbf{Germany}}  &  \multicolumn{3}{|c|}{\textbf{Spain}} \\
         & \textbf{EDGAR value}  & \textbf{Eurostat value} & \textbf{Percentage deviation} & \textbf{EDGAR value}  & \textbf{Eurostat value} & \textbf{Percentage deviation}\\
        \hline
         Industry & 188.81 & 115.80 & 38.67 & 81.24 & 37.86 & 53.40 \\ 
       \hline
         Buildings & 128.05 & 109.69 & 14.34 & 36.88 &  24.32 & 34.06 \\ 
         \hline
         Transport & 143.38 & 147.27 & -2.71 & 83.51 &  90.21 & -8.02 \\ 
         \hline
         Agriculture & 55.81 & 51.72 &  7,33 & 44.46 & 34.86 & 21.59\\ 
        \hline
    \end{NiceTabular}
    \caption{\label{tab:TV_EDGAR_sectoral_comparison} Comparison of sectoral values reported by EDGAR database and Eurostat values at NUTS0 level, to determine matching sectors. NOTE: The unit of measure of the values is kt CO2 equivalent.}
\end{table}

\end{document}